\documentclass[aps,prc,twocolumn, superscriptaddress,showkeys,nofootinbib]{revtex4-1}  

\usepackage[utf8]{inputenc}
\usepackage{graphicx}
\usepackage{subcaption}

\usepackage{booktabs}

\usepackage{graphicx,color}
\usepackage{amsmath,amssymb,amsfonts}
\usepackage{hyperref}

\usepackage{xspace}	
\input{defs.tex}

\begin{document}
\title{
Drell--Yan at the Electron-Ion Collider
}
\medskip
\author{Henry~T.~Klest} 
\email{hklest@anl.gov}
\affiliation{Physics Division, Argonne National Laboratory, \\
Lemont, IL 60439, USA}
\begin{abstract}
The photon is arguably the most universally important particle across all fields of physics. Despite its status as a fundamental particle, at high energies the photon can be seen as a hadronic source of partons. The partonic content of the photon is very poorly constrained compared to that of the proton, with photon PDF uncertainties typically one or two orders of magnitude larger than their proton counterparts, despite the fact that its source, the $\gamma\to q\bar{q}$ splitting, is perturbatively calculable. The high luminosity, excellent particle identification, and far-backward electron tagging capabilities of the Electron-Ion Collider make it an ideal environment for studying photon parton distribution functions. Similar to the $p+p$ or $\pi+p$ systems, photoproduction at the EIC can be thought of as two parton distributions colliding. One of the most powerful processes in such collisions is production of lepton pairs, i.e. $h+p\rightarrow l^+l^-+X$, known as the Drell--Yan process. This process has the ability to access for the first time the transverse-momentum-dependent parton distributions of the photon. The transversely polarized proton beam of the EIC additionally provides a possible means of accessing the transversity distribution of the proton without relying on fragmentation functions.

\end{abstract}
\maketitle

\section{Introduction} \label{sec:intro}
At high energies, the photon wavefunction contains a sizable hadronic component and therefore behaves as an effective source of partons~\cite{Kroll:1967it,Brodsky:1972vv,Walsh:1973mz,Bauer:1977iq}. The distribution of partons within this component of the photon wavefunction can be described via photon parton distribution functions (PDFs)~\cite{Worden:1974hc,Witten:1977ju,DeWitt:1978wn,Irving:1980qa}. The partonic structure of the photon is of particular interest since it is currently the only gauge boson for which the PDFs can be studied in experiments. The photon is a theoretically cleaner source of partons than any hadron; the pointlike coupling of the photon to quarks makes a substantial component of its PDFs calculable in perturbation theory from first principles. In spite of this calculability and nearly five decades of two-photon and photoproduction measurements, our understanding of the photon’s partonic structure remains an order of magnitude less precise than that of the proton. 

In contrast to the proton case, different fits of the photon PDFs display significant discrepancies in both size and flavor separation. Furthermore, no direct experimental constraints exist on the transverse-momentum-dependent (TMD) PDFs of the photon. Historically, the most precise information on the photon’s quark content has come from measurements of the inclusive photon structure function $F_2^{\gamma}(x,Q^2)$ in $e^+e^- \to e^+e^- + {\rm hadrons}$ with single- or double-tagged leptons~\cite{Brodsky:1971vm,ALEPH:2003F2gamma,AMY:1990F2gamma,AMY:1995F2gammaHighQ2,AMY:1997F2gamma,JADE1984_F2gamma,DELPHI:1996F2gamma,DELPHI2008_Dijet_gammagamma_EPJC57,L3_1999_2000_Q2Evolution_F2gamma,L3_2004_InclusiveJet_gammagamma_PLB602,OPAL:1994F2gamma,OPAL1997_Q2Evolution_arXiv,OPAL1997_ZPC74_AnalysisF2gamma,OPAL2000_Dstar_gammaGamma_EPJC16,OPAL:2000F2gammaLowX,OPAL2002_F2cgamma_PLB539,OPAL2003_Dijet_gammagamma_EPJC31,PLUTO:1981F2gamma,TASSO:1986F2gamma,TOPAZ:1994F2gamma,TPC2G_1987_PRL_F2gamma}. These measurements established the logarithmic $Q^2$ growth of $F_2^\gamma$ and the interplay between the pointlike and hadronic components of the photon\footnote{Similar $\gamma\gamma$ measurements can be performed on quasi-real photons generated by protons or ions at the EIC (See e.g. Ref.~\cite{Bertulani:2024vpt}), but this channel will not be the focus of this paper.}. The majority of photon PDF fits rely exclusively on data from $e^+e^-$ colliders~\cite{Gordon:1996pm,Cornet:2004nb,Gluck:1991jc,Schuler:1996fc}. For reviews of photon structure functions, see Refs.~\cite{Nisius:1999cv,Nisius:2009xx,Berger:2014rva}. 

Beyond inclusive structure functions, sensitivity to the gluon content of the photon has been inferred from heavy-flavor production and dijets in $\gamma\gamma$ collisions at LEP and in photoproduction at HERA. In $\gamma\gamma$ collisions, charm-tagged final states and dijet spectra have been used to access the resolved-photon component~\cite{OPAL2000_Dstar_gammaGamma_EPJC16,OPAL2002_F2cgamma_PLB539,OPAL2003_Dijet_gammagamma_EPJC31,DELPHI2008_Dijet_gammagamma_EPJC57,L3_2004_InclusiveJet_gammagamma_PLB602}. In $ep$ photoproduction, double-differential dijet cross sections and charged-particle spectra have been compared to different photon PDFs~\cite{H1_1998_EffectivePhotonPDF_EPJC1,H1_1999_ChargedParticles_GluonPhoton_EPJC10,H1_2000_VirtualPhotonPDF_EPJC13,ZEUS2002_DijetPhotoprod_arXiv0112029}. However, these $\gamma p$ data were generally not used in the fits of photon PDFs, with the exception of the SAL fit of Ref.~\cite{Slominski:2005bw}. It is important to note that no fit of the photon PDFs has been performed in a modern framework with uncertainty quantification.

\begin{figure}[h]
    \centering
    \qquad 
    \includegraphics[width=0.9\linewidth,
        trim={0.1cm 0.1cm 0.1cm 0.1cm},
        clip
    ]{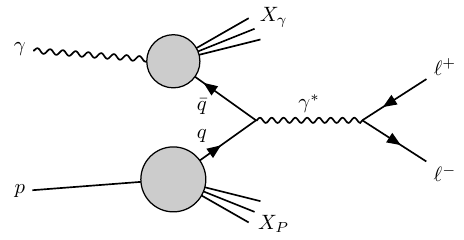}
    \caption{Example diagram of the leading-order Drell--Yan process in photoproduction.}
    \label{fig:Feynman}
\end{figure}

The Electron--Ion Collider~\cite{AbdulKhalek:2021gbh} (EIC) offers a qualitatively new opportunity to address these questions by measuring for the first time the Drell--Yan process in photoproduction, i.e.,
\begin{equation}
\gamma p\rightarrow\ell^+\ell^-+X,
\end{equation}
where the initial-state photon is supplied by the electron beam. A leading-order diagram contributing to this process is sketched in Fig.~\ref{fig:Feynman}, where $X_{\gamma}$ and $X_p$ represent the hadronic remnants of the photon and proton, respectively. The power of measurements of di-jet photoproduction at the EIC to constrain the polarized and unpolarized PDFs of the photon has been demonstrated in Ref.~\cite{Chu:2017mnm}. The hadron-beam Drell--Yan process, $h+h\rightarrow\ell^+\ell^-+X$, is a primary tool in the phenomenology of both collinear and TMD parton distributions\footnote{``Drell--Yan" is often used colloquially to refer to the quark-antiquark annihilation subprocess, but other subprocesses such as QCD Compton scattering ($q+g\rightarrow q+\gamma^*$) also produce virtual photons which decay to a dilepton pair. The subprocess which produced the virtual photon cannot be distinguished experimentally, so the various subprocesses need to be calculated and included at the level of phenomenological analyses. For the purposes of this paper we will refer to all QCD virtual-photon-producing inelastic processes as Drell--Yan.}. For reviews, see Refs.~\cite{Reimer:2007Review,Peng:2014hta}. In addition to $p+p$ collisions, experimental measurements of $\pi/K+p\rightarrow\ell^+\ell^-+X$ on polarized and unpolarized proton targets have provided novel constraints on the partonic structure of both beam and target~\cite{COMPASS2017DYTSA,COMPASS:2024PRL,Peng:2017KaonJpsi,Bourrely:2024KaonPDF,E615:1989PRD,Barry:2018PRL,Novikov:2020xFitter,NA10:1988ZPC}. 

In the \emph{resolved-photon} photoproduction channel, the leading-order hard subprocess $q \,\bar q \to \gamma^\ast \to \ell^+\ell^-$ provides access to the PDFs of the photon and proton at scales $Q=M_{\ell\ell}$, where $M_{\ell\ell}$ is the invariant mass of the dilepton pair. Together with $M_{\ell\ell}$, the rapidity of the dilepton pair $y_{\ell\ell}$ constrains the values of Bjorken $x$ of the photon and the proton. At leading order in the $\gamma p$ center-of-mass frame\footnote{We choose the convention where the electron and the quasi-real photon travel along the $-\hat{z}$ direction and the proton travels in the $+\hat{z}$ direction, in line with the typical EIC convention where the proton moves in the $+\hat{z}$ direction. We neglect any transverse momentum of the initial-state photon.},
\begin{equation}
x_\gamma \simeq \frac{M_{\ell\ell}}{\sqrt{s_{\gamma p}}} e^{-y_{\ell\ell}},\qquad
x_p \simeq \frac{M_{\ell\ell}}{\sqrt{s_{\gamma p}}} e^{+y_{\ell\ell}},
\end{equation}
with $s_{\gamma p}\simeq y\,s_{ep}$ set by the photon’s inelasticity $y$. In photoproduction at $Q^2=0$, $y$ is equal to $E_\gamma/E_e$~\footnote{There is an unfortunate degeneracy in notation between the inelasticity $y$ and the dilepton rapidity $y_{\ell\ell}$. Wherever $y$ appears without a subscript, it should be taken as the inelasticity; rapidity will always appear with a subscript.}. Thus, differential measurements in $M_{\ell\ell}$ and $y_{\ell\ell}$ map out the phase space of $x_\gamma$ and $x_p$. Resolved photoproduction corresponds to the kinematic region of $x_\gamma<0.8$ or so. Drell--Yan in the proton valence region will predominantly select antiquarks from the photon, giving the first independent handle on the photon’s antiquark PDFs. 

The direct photoproduction channel, where the photon interacts directly with a quark as a pointlike object, corresponds to $x_{\gamma}\approx1$. Direct photoproduction of dilepton pairs is also of interest for proton structure studies, in part because of the similarity of the process to timelike Compton scattering (TCS)~\cite{BrennerMariotto:2013dal,Machado:2008zv}. Interference between Bethe--Heitler and partonic amplitudes offers an additional, qualitatively different handle on photon and proton structure. In deeply virtual Compton scattering (DVCS) and TCS, the interference of Bethe--Heitler with the Compton amplitude is routinely exploited to gain access to the generalized parton distributions of the proton. An analogous strategy can be pursued in dilepton photoproduction. In the two--jet topology where Bethe--Heitler and direct Drell--Yan become experimentally indistinguishable, suitably chosen lepton--angular asymmetries can isolate the interference between Drell--Yan and Bethe--Heitler. This interference is proportional to a novel combination of quark flavors and thus provides additional sensitivity to the photon and proton parton distributions beyond what could be gained with Drell--Yan alone. The direct photoproduction lepton-charge asymmetry is thus complementary to both inclusive DIS measurements and traditional hadron beam Drell--Yan (see Sec.~\ref{sec:DYBH}). Furthermore, the interference of Bethe--Heitler and Drell--Yan with an additional chiral--odd subprocess accessible in the case of a transversely polarized target provides a novel avenue for the EIC to access the nucleon transversity distribution. This means of studying transversity is complementary to semi-inclusive DIS~\cite{Pire:2009ap} and does not rely on the use of fragmentation functions. The high polarization of the EIC beams make it the ideal facility to perform such a measurement for the first time. These aspects of the direct photoproduction channel are explored in Sec.~\ref{sec:DYBH}.

The Drell--Yan photoproduction process was first discussed in the literature by Jaffe\footnote{Although the author of Ref.~\cite{Jaffe:1971we} thanks Professor Drell for suggesting the problem.} in 1971~\cite{Jaffe:1971we} as a process to test the parton model in experiments at SLAC. The process has been subsequently discussed theoretically and phenomenologically in the context of various experimental facilities in Refs.~\cite{Kang:1979xe,Jones:1979wa,Busenitz:1981hr,Badalian:1990eq,Bawa:1993qr,Bussey:1996vq,Pire:2009ap,BrennerMariotto:2013dal}\footnote{The 1981 paper of Busenitz and Sullivan, Ref.~\cite{Busenitz:1981hr}, optimistically describes the cross section as ``depressingly small" and measurement of this process as ``very difficult and perhaps impossible."}. This process is in principle also accessible in ultra-peripheral collisions at hadron colliders~\cite{Jones:1979wa}. In spite of over 50 years of history, the Drell--Yan photoproduction process has never been measured in an experiment. However, the combination of high luminosity, large acceptance, low-$Q^2$ tagging, variable beam energies, target polarization, and particle identification make the EIC an optimal environment for not only measuring this process for the first time but also multi-differentially mapping it in detail. The unique ability of the EIC to measure a jet in the photon-going hemisphere further enhances the resolved component and enables jet flavor tagging to aid in separation of quark species in the photon, as demonstrated nicely in Ref.~\cite{Chu:2017mnm}.

Drell--Yan at the EIC furthermore has the potential to provide the first constraints on \emph{photon TMDs}. In addition to the dilepton mass and rapidity, the transverse momentum, $p_{T,\ell\ell}$, and the Collins--Soper angular coefficients can be reliably measured at the EIC. These variables provide access to the intrinsic transverse motion of partons in the beams and, assuming the proton TMDs are known well from SIDIS measurements at the EIC, can be used to infer the TMDs of the photon. The high luminosity and large detector acceptance enable differential mapping of $x_\gamma$ and $p_{T,\ell\ell}$.

One of the most interesting results from $\pi+p$ Drell--Yan with polarized protons is the recent observation of the sign flip in the Sivers effect compared to semi-inclusive DIS~\cite{COMPASS:2023vqt,Peng:2025etb}. This process provides a direct test of the predicted process dependence of TMDPDFs, driven by the color flow encoded in initial- and final-state interactions. In particular, the Sivers function is expected to change sign between SIDIS and Drell–Yan, reflecting the nontrivial gauge-link structure of QCD factorization for TMDs. In this work we assess the capabilities of the EIC to measure Drell--Yan target–spin asymmetries. Such measurements would provide the unique opportunity to observe the predicted sign change of the Sivers function between SIDIS and Drell--Yan within a single collider experiment and using the same polarized proton beam.

In this paper, we develop a concrete measurement strategy for the EIC based on simulations of the Drell--Yan channel in the new resolved photoproduction framework in the \textsc{Sherpa} event generator~\cite{Hoeche:2023gme}. Recent progress in producing precise predictions for resolved photoproduction with \textsc{Sherpa}~\cite{Hoeche:2023gme,Meinzinger:2023xuf}, Pythia~\cite{Helenius:2017aqz}, and MadGraph~\cite{Manna:2024ltm} has made it clear that the photon PDF is now the dominant source of uncertainty in photoproduction cross section predictions~\cite{Meinzinger:2023xuf}. Precise Monte Carlo predictions, including the input PDFs, are necessary to enable the experimental precision expected from the EIC. High-$y$ inclusive and semi-inclusive measurements, including measurements targeting gluon saturation, are particularly sensitive to the modeling of photoproduction backgrounds. The existing photon PDF sets predict dramatically different cross sections for resolved photoproduction, which can translate into non-negligible uncertainties on measured cross sections arising from the lack of knowledge of the background. Therefore, in addition to being of intrinsic interest, a modern re-fit of the photon PDFs with more precise data would help improve the precision of many key EIC measurements.


\section{Analysis \& Simulation}\label{sec:strategy}

\subsection{Simulation}
We simulate the signal process of $\gamma p\rightarrow l^+l^-+X$ for the EIC configuration of 10 GeV electrons on 275 GeV protons (henceforth denoted as 10x275) using \textsc{Sherpa} 3.0.1~\cite{Sherpa:2019gpd} with the NLO resolved photoproduction model described in Ref.~\cite{Hoeche:2023gme}. The 10x275 beam energy configuration is chosen because it offers the highest luminosity. The design luminosity for the 10x275 configuration is 100 fb$^{-1}$ per year, around a factor of 7 higher than 18x275. In the \textsc{Sherpa} simulation, the initial-state photon is sourced from the electron via the Weizsäcker-Williams equivalent photon approximation. $Q^2_{\mathrm{max}}$ is set to 1 GeV$^2$ and the initial-state photons are assumed to be collinear to the electron beam. The matrix elements for the hard scattering process $q\bar{q}\to\ell^+\ell^-$ are handled by Amegic~\cite{Krauss:2001iv}, Comix~\cite{Gleisberg:2008fv,Hoeche:2011fd}, and OpenLoops~\cite{Cascioli:2011va}. The direct photoproduction contribution to dilepton production is not included. The parton distributions of the photon are the SaSg PDFs of Refs. \cite{Schuler:1995fk} and \cite{Schuler:1996fc}, while the proton PDF is NNPDF 3.1~\cite{Ball:2014uwa} at NNLO and the strong coupling constant was set to $\alpha_s=0.118$. Multi-parton interactions, which have been observed to play a non-negligible role in photoproduction~\cite{ZEUS:2021qzg,Butterworth:2024hvb}, are included in the \textsc{Sherpa} model. Additional information on the various components of the simulations can be found in Refs.~\cite{Bothmann:2016nao,Buckley:2019xhk,Schonherr:2008av,Chahal:2022rid,Denner:2002ii,Denner:2005nn,Denner:2010tr,Schumann:2007mg,Jadach:1999vf}.

Based on these ingredients, \textsc{Sherpa} is able to numerically evaluate the cross section and perform scale variations to understand the uncertainty on the prediction. In Ref.~\cite{Hoeche:2023gme}, the \textsc{Sherpa} authors demonstrate the good agreement between their predictions and the HERA data as well as the large uncertainty engendered by the photon PDF.

For production of Drell--Yan dimuon pairs at the 10x275 beam energy with $M_{\ell\ell}>1$ GeV, \textsc{Sherpa} with the SAS1M photon PDF set predicts a total cross section of 5.43$^{+1.34}_{-1.25}$ pb, while the same setup with the SAS2M photon PDF predicts a total cross section of 8.10$^{+0.47}_{-1.13}$ pb. The uncertainties provided on the \textsc{Sherpa} cross sections are the extrema of a 7-point scale variation. SAS2M was shown to be a better fit to world data on $F_2^\gamma$ in Ref.~\cite{Nisius:1999cv}, producing a $\chi^2/\mathrm{ndf}=1.01$ compared to the $\chi^2/\mathrm{ndf}=1.50$ of SAS1M.  The significant difference in predicted cross sections between the two PDF sets highlights the large uncertainty in the photon PDF, as well as the high constraining power of the Drell--Yan process. The SAS1M PDF set employs a starting scale of $Q_0^2=0.36$ GeV$^2$, while SAS2M uses a starting scale of $Q_0^2=4$ GeV$^2$. Both sets use the $\overline{\mathrm{MS}}$ renormalization scheme~\cite{Nisius:1999cv,Schuler:1996fc,Schuler:1995fk}. The discrepancy between the two sets does not go away at higher $Q^2=M_{\ell\ell}^2$, as can be seen from Fig.~\ref{fig:EventYields}. The differences in the derived photon structure function $F_2^{\gamma}$ are around 30\% between the two PDF sets~\cite{Nisius:1999cv}. These cross sections are more or less consistent with other estimates in the literature, namely those of Refs.~\cite{Jaffe:1971we,Jones:1979wa,Bussey:1996vq,Bawa:1993qr,BrennerMariotto:2013dal}. The cross section for the 18x275 beam energy in \textsc{Sherpa} is around 50\% higher than for 10x275.



\subsection{Kinematic Reconstruction}\label{subsec:yReco}

The techniques available for reconstruction of the kinematics in photoproduction depend on whether or not the scattered electron is measured. For tagged photoproduction where the electron is measured, the standard inclusive kinematic reconstruction methods can be employed (see Ref.~\cite{Bassler:1994uq} for a succinct review). 

\begin{figure}[h]
    \centering
    \qquad 
    \includegraphics[width=1.0\linewidth,
        trim={0.1cm 0.1cm 0.1cm 0.1cm},
        clip
    ]{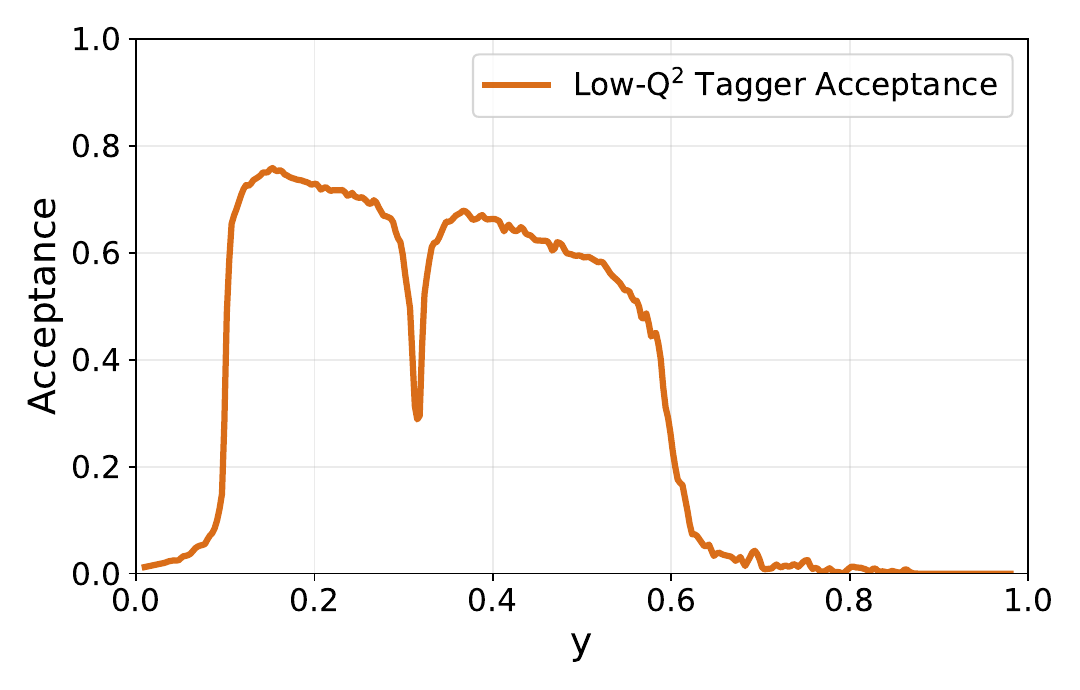}
    \caption{Acceptance of the ePIC low-$Q^2$ tagger as a function of inelasticity $y=E_\gamma/E_e$. The polar angle of the scattered electron is assumed to be negligible.}
    \label{fig:LowQAcceptance}
\end{figure}

The low-$Q^2$ tagger in the Electron-Proton/Ion Collider (ePIC) experiment has excellent resolution on the kinematics of the scattered electron, providing sub-percent resolution on $y$~\cite{Gardner:2023lly}. However, the acceptance is limited in $y$ and $\theta_{e'}$. The acceptance as a function of $y$ is shown in Fig.~\ref{fig:LowQAcceptance}. When the low-$Q^2$ tagger is required, the photon energy is henceforth restricted to the range of 1-4 GeV for the 10 GeV electron beam configuration. A reasonable strategy to extend the kinematic reach to larger $y$ would be to perform tagged measurements where available and use them to benchmark and understand the biases of untagged samples. 

Another generic challenge is the very high rate of background electrons striking the Low-$Q^2$ tagger, resulting in pileup where more than one electron is detected per event. The vast majority of these background electrons come from radiative elastic scattering, which peaks at very small $Q^2$, meaning a modest cut of $Q^2>10^{-3}$ GeV$^2$ can significantly improve the signal-to-background for scattered electrons. For the semi-inclusive event topologies relevant for this study, the pileup can be further mitigated by a matching of the event kinematics of the scattered electron and the hadronic final state measured in the central detector.

To study detector effects, we implement detector smearing with a set of resolutions and acceptances approximating the performance of ePIC. In our tagged sample, we simulate the acceptance by randomly accepting events according to the acceptance of the low-$Q^2$ tagger, shown in Fig.~\ref{fig:LowQAcceptance}. Our generated dataset is integrated over $Q^2$, so we also integrate the low-$Q^2$ tagger acceptance to get an acceptance purely as a function of $y$. 

In untagged samples, $E'_e$ and $\theta'_e$ are unknown, so $y$ must be reconstructed via
\begin{align}
y_{h} &\equiv \frac{\sum_{h}(E-p_z)}{E_e}, \label{eq:yh}
\end{align}
where the sum runs over all reconstructed particles except for the scattered electron. In the resolution and acceptance situation described in Appendix~\ref{sec:appendix}, the estimated resolution of $y_h$ is $\sigma(y)\simeq 0.18$ (roughly independent of $y$) with a non-negligible bias driven by losses of $E-p_z$ in the electron-going direction. For $y<0.2$, this translates into a very poor $>100\%$ resolution on $s_{\gamma p}$. This result is to be expected since at low-$y$ the photon has very little energy, resulting in a center-of-mass that is highly boosted in the proton-going direction and thus more susceptible to acceptance losses. Uncertainties in $y$ propagate directly to $x_\gamma$ and $x_p$ because $\sqrt{y\,s_{ep}}$ enters directly into the calculation of $x_{\gamma,p}$. Consequently, for untagged photoproduction, it seems unlikely that ePIC can bin in more than two or three bins in $x_{\gamma,p}$ due to poor resolution. We therefore focus on tagged photoproduction for the remainder of this paper.

One unique aspect of this program at the EIC is the ability to identify hadron species in jets. For resolved-photon events, which nominally arise from a $q\bar{q}$ configuration of the photon, a parton of a given flavor participates in the hard scattering, while the partner of that parton forms the photon remnant. Identifying the hadron species inside the remnant jet in the photon hemisphere can provide flavor information on the struck parton in the photon PDF. The capabilities of jet flavor identification in photoproduction have already been explored in considerable detail in Ref.~\cite{Chu:2017mnm}, to which we refer the reader for a more comprehensive discussion. This method of PDF flavor separation is also conceptually similar to the HERA measurements of heavy flavor di-jets in resolved photoproduction~\cite{H1:2012oud,ZEUS:2011aa}. In the $\gamma p$ center-of-mass frame, the remnant is emitted predominantly into the photon hemisphere, i.e., $y^\ast\!\ll\!0$, where $y^\ast$ refers to the rapidity in the $\gamma p$ center-of-mass frame as opposed to the lab frame. The amount of energy available for the photon remnant is $E_{\gamma}(1-x_{\gamma})$. Thus, if $E_{\gamma}$ is known from the tagged scattered electron, the remnant jet energy can therefore be used as an alternative method to reconstruct $x_{\gamma}$. One limitation is that the momentum available for the remnant can become small when $y$ is small and $x_{\gamma}$ is large. Most likely a special jet reconstruction algorithm which takes into account the kinematic information of the event (à la Centauro~\cite{Arratia:2020ssx}) will be required to reconstruct jets in this case. After applying detector effects, the resolution on the momentum of the hadronic final state in the remnant hemisphere is around 300 MeV and is roughly constant with energy. Therefore, at small $y$ and large $x_{\gamma}$ the resolution on $x_{\gamma}$ from the remnant method degrades significantly.

Using the expected tracking resolutions of the ePIC detector, the resolutions on the lepton pair mass, transverse momentum, and $\gamma p$ center-of-mass frame rapidity are shown in Fig.~\ref{fig:EventYields}. The final product of the analysis would be an inclusive dilepton spectrum similar to the one shown in Fig.~\ref{fig:DY_Mll}, or in Refs.~\cite{PHENIX:2009gyd,PHENIX:2018dwt}, from which the Drell--Yan contribution can be extracted after a careful background subtraction. The response matrix for $x_{\gamma}^{\mathrm{LO}}=\frac{M_{\ell\ell}}{\sqrt{s_{\gamma p}}} e^{-y_{\ell\ell}}$ in \emph{tagged} photoproduction is shown in Fig.~\ref{fig:xGammaRes}. The resolution on $x_{\gamma}^{\mathrm{LO}}$ is somewhat better in Drell-Yan than in the previous results for dijet production (See Fig. 4 of Ref.~\cite{Chu:2017mnm}). This improvement arises from the fact that the Low-$Q^2$ tagger is now a tracking system with better resolution than the calorimeter that was assumed in Ref.~\cite{Chu:2017mnm} and because the kinematic reconstruction in Drell-Yan relies only on single charged particles, which are generally less susceptible to large resolution fluctuations than jets.

Triggering is a non-trivial aspect of this measurement because the signal is embedded in an enormous rate of inclusive photoproduction events. At HERA, photoproduction triggers were therefore typically heavily prescaled. For the dielectron channel, $\gamma p\to e^+e^-+X$, the signature of an $e^+e^-$ pair in the central detector should be relatively easy to trigger on, even for a more traditional trigger system. Dimuon selections, however, were practical at HERA only because dedicated muon systems provided a relatively clean low-level signal to trigger on, but such muon systems are currently not foreseen for ePIC. At the EIC, the intent is to use streaming readout with no hardware trigger, so the keep/reject decision can be deferred until after more advanced online reconstruction and feature building (e.g., cluster-track matching) have been performed~\cite{AbdulKhalek:2021gbh,Ameli:2022gvw,Brei:2025isx}. Requiring two tracks pointing to MIP-like energy deposits throughout the calorimeters will significantly reduce the background rate. If a hadron rejection factor of 50 can be obtained online for each muon, it would reduce the rate of events being read out by a factor of 2500 compared to a simple trigger on two tracks. This means that streaming readout should allow the dimuon channel to be retained with higher efficiency even in inclusive photoproduction, where a traditional hardware trigger would likely be forced into aggressive prescales or thresholds. For the remainder of this paper, we assume that the topology of $e'$ in the low-$Q^2$ tagger and $\ell^+\ell^-$ in the central detector can be accepted with good efficiency. In reality, the feasibility of accepting the dimuon channel will ultimately depend on the final implementation of the data acquisition system.

\begin{figure*}[ht]
    \centering
    \begin{subfigure}[t]{0.95\linewidth}
        \centering
        \includegraphics[width=\linewidth,
            trim={0.1cm 0.1cm 0.1cm 0.1cm},
            clip
        ]{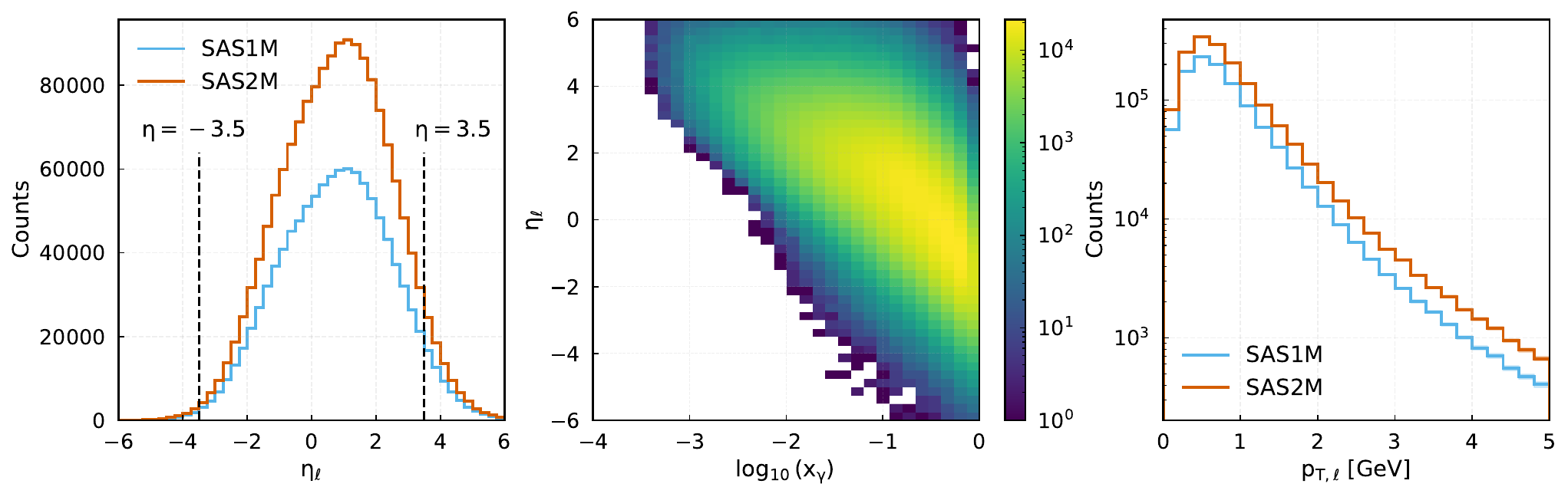}
    \end{subfigure}
    \caption{\textbf{Left:} Pseudorapidity distributions of leptons produced in the Drell--Yan process for the 10x275 GeV configuration, as predicted by \textsc{Sherpa} with the SAS2M and SAS1M PDF sets. The nominal ePIC acceptance of $|\eta|\leq3.5$ is the region between the black dashed lines. \textbf{Center:} Distribution of lepton pseudorapidity as a function of the true $x_{\gamma}$. \textbf{Right:} Lab frame $p_T$ distributions of Drell--Yan leptons.}
           \label{fig:DY_LabFrame}
\end{figure*}

\begin{figure}[h]
    \centering
    \qquad 
    \includegraphics[width=1.0\linewidth,
        trim={0.1cm 0.1cm 0.1cm 0.1cm},
        clip
    ]{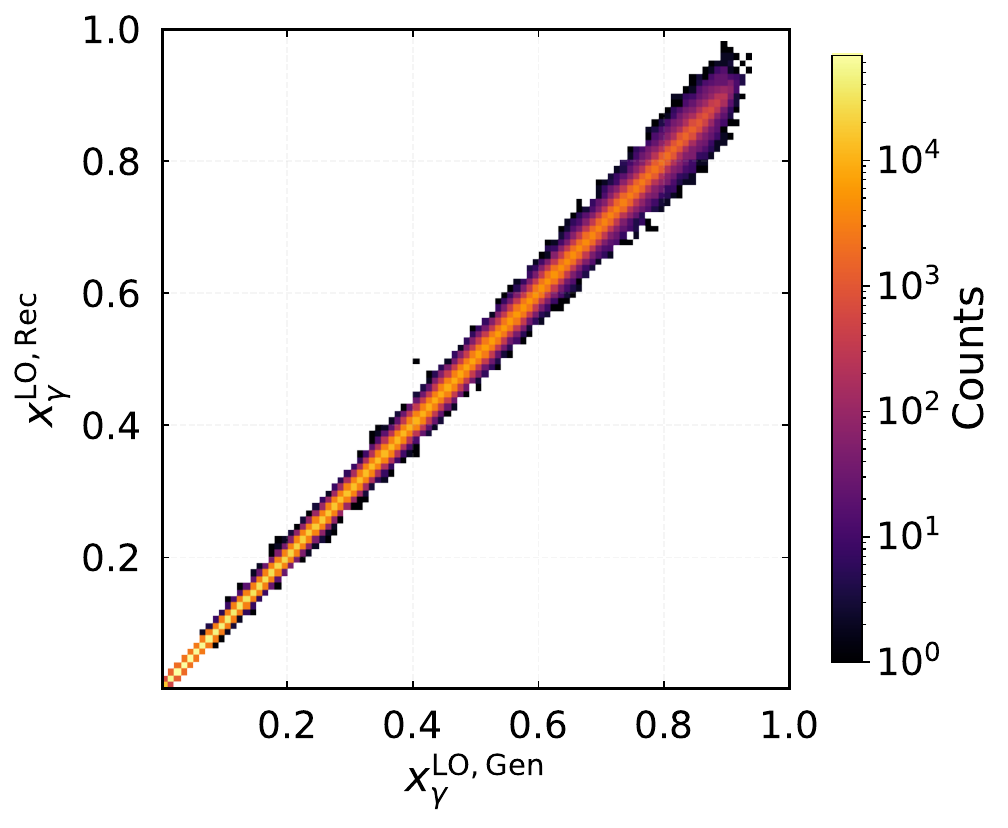}
    \caption{Response matrix of $x_{\gamma}^{\mathrm{LO}}=\frac{M_{\ell\ell}}{\sqrt{s_{\gamma p}}} e^{-y_{\ell\ell}}$, assuming the experimental acceptances and resolutions discussed in Appendix~\ref{sec:appendix}. Note the logarithmic color scale.}
    \label{fig:xGammaRes}
\end{figure}

\subsection{Backgrounds}\label{subsec:Backgrounds}

The largest physics background for Drell--Yan at relatively low masses ($M_{\ell\ell}<4$ GeV) arises from decays of hadrons to leptons~\cite{PHENIX:2009gyd,PHENIX:2018dwt}. For this reason, experiments targeting the Drell--Yan process historically have considered masses greater than the $J/\psi$ mass. However, the excellent tracking resolution of the EIC means that peaking sources of dileptons such as the $J/\psi$ should be very narrow and can be subtracted from the continuum. Measurements of low-mass dielectrons and dimuons have been performed in $p+p$ collisions for example by PHENIX in Refs.~\cite{PHENIX:2009gyd,PHENIX:2018dwt}, where it was observed that in the mass range between the $\phi$ and $J/\psi$ mesons, the $M_{\ell\ell}$ spectrum was dominated by semi-leptonic decays of charmed hadrons arising from correlated $c\bar{c}$ pairs. This background, in principle, can be suppressed at the EIC with precise vertexing. The removal of displaced vertices from semi-leptonic weak decays of $D$-mesons and other charm hadrons can help reduce the background sources of dilepton pairs in the continuum region of 1-3 GeV.

\begin{figure}[h]
    \centering
    \begin{subfigure}[t]{0.49\linewidth}
        \centering
        \includegraphics[width=\linewidth,
            trim={0.1cm 0.1cm 0.1cm 0.1cm},
            clip
        ]{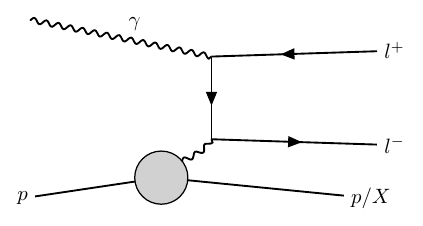}
        \label{fig:FeynmanBkgd}
    \end{subfigure}
    \hfill
    \begin{subfigure}[t]{0.49\linewidth}
        \centering
        \includegraphics[width=\linewidth,
            trim={0.1cm 0.1cm 0.1cm 0.1cm},
            clip
        ]{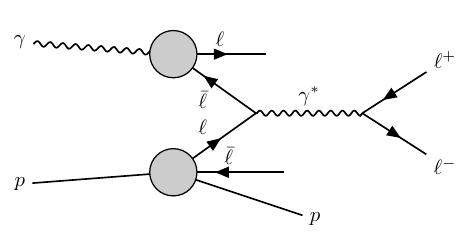}
        \label{fig:FeynmanQED}
    \end{subfigure}
    \caption{Example diagrams for QED processes contributing to inclusive dilepton production. \textbf{Left:} Bethe--Heitler process. \textbf{Right:} QED Drell--Yan Process where an antilepton from the photon annihilates with a lepton from the proton PDF.}
    \label{fig:FeynmanProcesses}
\end{figure}

\begin{figure*}[ht]
    \centering
    \begin{subfigure}[t]{0.95\linewidth}
        \centering
        \includegraphics[width=\linewidth,
            trim={0.1cm 0.1cm 0.1cm 0.1cm},
            clip
        ]{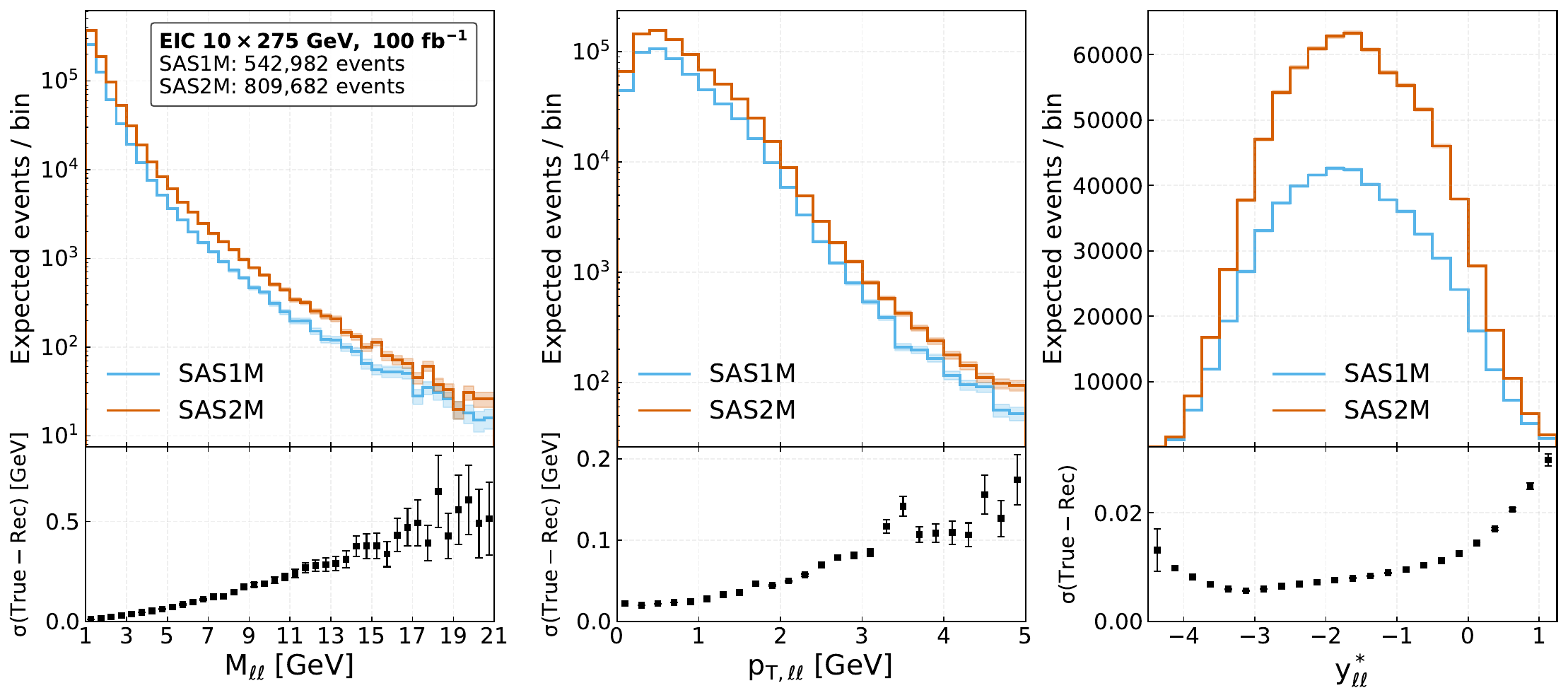}
    \end{subfigure}
    \caption{Expected reconstructed event yields for \textit{tagged} photoproduction at the EIC for the SAS1M and SAS2M photon PDF sets. The yields are shown for a single dilepton channel, i.e., they would be doubled if including both dielectron and dimuon channels. Detector acceptances and resolutions are imposed on the scattered electron and dilepton pair. The lower panels provide the expected detector resolutions on the variables given the detector parameters in Appendix~\ref{sec:appendix}.}
    \label{fig:EventYields}
\end{figure*}

There are also various physics processes which produce the same continuum dilepton final state. One of these is the Bethe--Heitler reaction shown in the top panel of Fig.~\ref{fig:FeynmanProcesses}, whereby the incoming photon interacts with a photon originating from the proton and produces a lepton pair~\cite{Chwastowski:2022fzk}. One can also imagine a dilepton background contribution from the process where the electron scatters on a positively charged lepton from the lepton PDF of the proton, as discussed in Ref.~\cite{DaRold:2024ram}. The primary background to Drell-Yan arises from the case where the photon splits into a dilepton pair, and one of the leptons emits a virtual photon which scatters inelastically on a parton in the proton and produces two jets: a target remnant and a current jet. This process is shown in the top left of Fig.~\ref{sec:DYBH}. The elastic or quasi-elastic case where the proton remains intact or dissociates into a low-mass state can be easily distinguished and subtracted via the pattern of produced final-state hadrons. Although the Drell--Yan and inelastic Bethe--Heitler amplitudes formally interfere (a potentially useful feature that will be discussed in Sec.~\ref{sec:DYBH}), the two contributions populate rather different regions of phase space, so that one can select kinematics where the cross section is dominated by Drell--Yan and the Bethe--Heitler background is small~\cite{Busenitz:1981hr}. Bethe--Heitler events reside almost entirely at $x_\gamma \simeq 1$ and very small $|t|$, with the lepton pair boosted into the photon hemisphere, whereas resolved-photon Drell--Yan populates the full region $x_\gamma<1$ and has a much flatter rapidity dependence~\cite{Busenitz:1981hr}. By requiring $x_\gamma<1$ and moderate or negative dilepton rapidity, the Bethe--Heitler contribution can be driven below the percent-level relative to Drell--Yan, as observed in Ref.~\cite{Busenitz:1981hr}. The angular distributions of the leptons in the dilepton rest frame can further discriminate Drell--Yan from purely QED production~\cite{Bussey:1996vq}, as the Bethe--Heitler contribution peaks very strongly at certain angles while Drell--Yan does not. The remaining contribution of Bethe--Heitler can be calculated and subtracted as a background.

In Drell--Yan measurements, often the dimuon final state is the channel of choice for experimental measurements. At the EIC, both the dielectron and dimuon channels are available, but each has their own unique challenges. For dimuons, the challenge is cleanly identifying muons amongst a large pion background. The yield of pions in minimum bias $ep$ events is approximately a factor of 500 higher than the yield of muons. The desired pion rejection factor to reconstruct dimuon events with 90\% efficiency and 10\% contamination is around 5000. Unlike most fixed-target Drell--Yan or $pp$ collider experiments, ePIC does not have dedicated muon identification detectors. A cut on the energy deposited in the calorimeters can provide a pion rejection factor of about 100, which is sufficient for reconstructing vector mesons whose yields can be extracted from invariant-mass peaks above background, but for continuum processes such as Drell--Yan, a higher muon purity is needed. Depending on the muon momentum, the time-of-flight and Cherenkov PID systems will provide additional muon--pion separation; however, the combined performance is unlikely to reach the level required for a clean dimuon measurement\footnote{Comprehensive studies of the muon identification capabilities of ePIC are currently ongoing.}. Such a measurement may therefore be better suited to a second EIC detector optimized for high-purity muon detection~\cite{Jacobs:2025rqa}. 

For the dielectron channel, a DIS scattered electron, particularly if it radiates a photon in the initial-state, can be misidentified as a Drell--Yan pair electron and erroneously paired with a positron from a hadronic decay. Since the DIS cross section is much larger than the Drell--Yan cross section and also will produce a continuum shape, this background is an added complication for the dielectron measurement. With the inclusion of tagging the scattered electron in the Low-$Q^2$ tagger, however, this background should be reduced. In contrast to the dimuon channel, the purity of electrons in ePIC will be very high, particularly in the region of $\eta<1.4$ corresponding to the barrel and backward regions where the combined hadron rejection power of particle identification and calorimetry is expected to be on the order of 10$^4$. Around two-thirds of the Drell--Yan leptons have $\eta<1.4$, and this region generally corresponds to larger values of $x_{\gamma}$. Another downside of the dielectron channel is the increased magnitude of radiative corrections.

One unique opportunity at the EIC for relatively large $M_{\ell\ell}$ is to exploit the $\gamma^*\rightarrow\tau^+\tau^-$ channel. Reconstruction of $\tau$ leptons has been discussed at the EIC in the context of charged-lepton flavor violation and measuring the $\tau$ anomalous electromagnetic moment~\cite{Zhang:2023EICetau,Deng:2025hio,Chwastowski:2022fzk}. While the number of usable di-tau events will be significantly lower than dielectrons or dimuons due to unfavorable branching fractions, the di-tau channel provides a dramatically reduced background due to the displaced vertex of the $\tau$. The combination of three dilepton final-states, each of which have different sources of background, will provide a cross-check and improve the understanding of the Drell--Yan component of the inclusive dilepton spectrum.

\section{Results}
\subsection{Cross Sections}
The expected top luminosity of the EIC, which is available only at the 10x275 GeV beam energy, is 100 fb$^{-1}$ per year. In one year of running, the expected number of Drell--Yan dileptons with $M_{\ell\ell}>1$ GeV is on the order of one to two million for the dielectron and dimuon channels combined. For the ``safe" Drell--Yan region of $M_{\ell\ell}>4$ GeV, we expect 50,000 to 100,000 tagged and reconstructed Drell--Yan pairs in 100 fb$^{-1}$. This number will necessarily be further reduced by analysis cuts to remove, for example, pileup in the low-$Q^2$ tagger or backgrounds to the final-state leptons. We forego estimating systematic uncertainties at this stage since they will depend on the understanding of the detector and background, which will evolve over the lifetime of the experiment.


One of the fortuitous observations is that for the 10x275 GeV beam energy configuration, the majority of the Drell--Yan dilepton pairs enter the central detector acceptance, as can be seen from Fig.~\ref{fig:DY_LabFrame}. This occurs because the photon PDF prefers larger values of $x$ than the proton PDF, but the energy of the proton beam is much higher than the photon. The result is that the quark-antiquark collisions are not dramatically boosted in either the proton-going or electron-going directions in the lab-frame. The center panel of Fig.~\ref{fig:DY_LabFrame} shows that the detector acceptance in the forward direction does not significantly limit the reach to large $x_{\gamma}$, but the reach to low-$x_{\gamma}$ is somewhat limited by the acceptance in the backward direction.

\begin{figure}[h]
    \centering
    \qquad 
    \includegraphics[width=1.0\linewidth,
        trim={0.1cm 0.1cm 0.1cm 0.1cm},
        clip
    ]{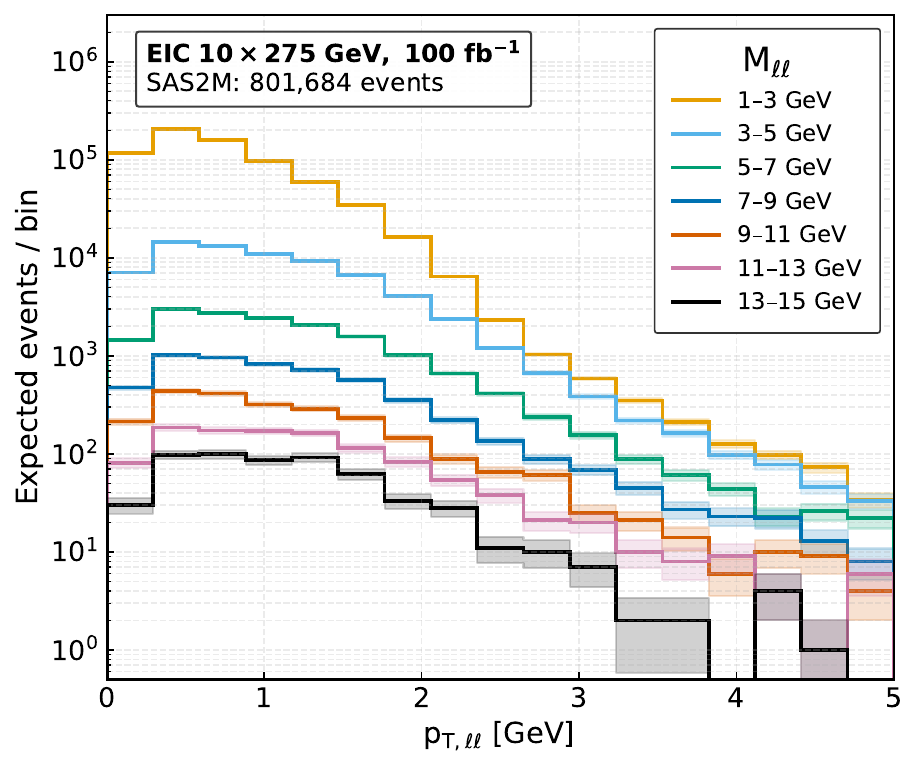}
    \caption{Example $p_{T,\ell\ell}$ distributions in bins of $M_{\ell\ell}$. Uncertainties are statistical only.}
    \label{fig:DY_pT}
\end{figure}

\begin{figure}[h]
    \centering
    \qquad 
    \includegraphics[width=1.0\linewidth,
        trim={0.1cm 0.1cm 0.1cm 0.1cm},
        clip
    ]{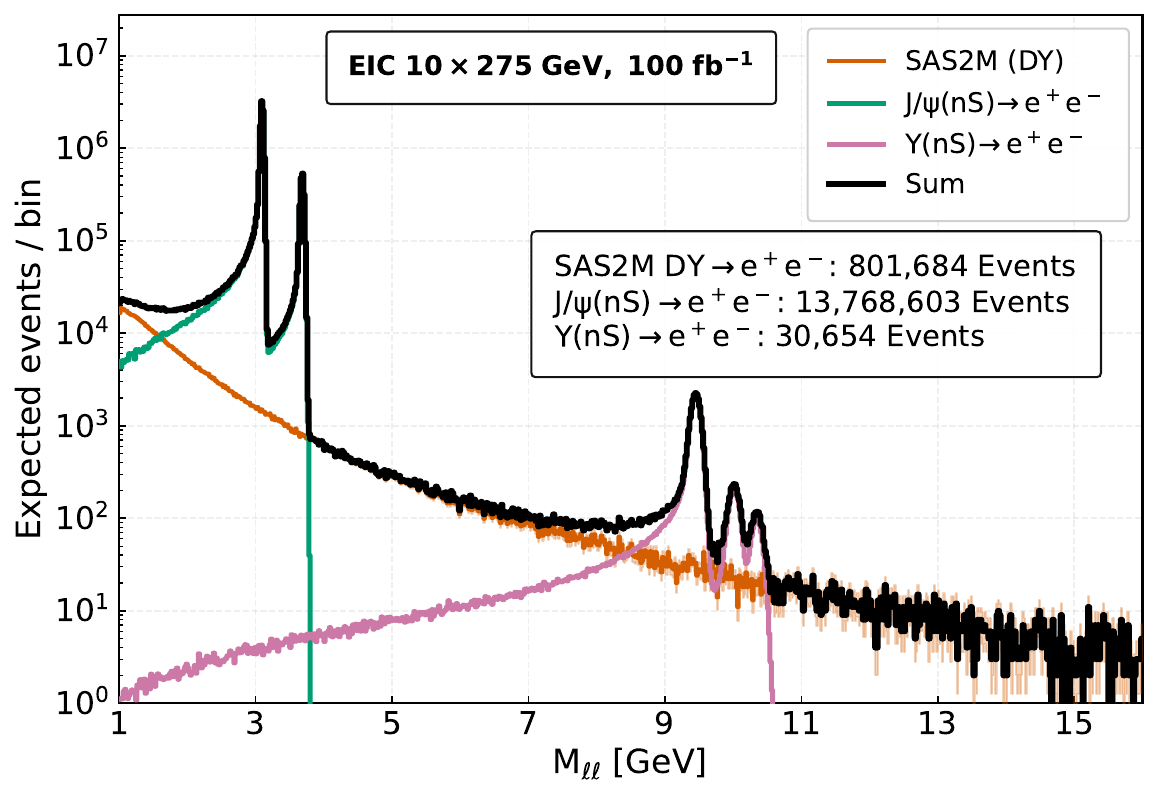}
    \caption{Example reconstructed dilepton invariant mass spectrum from Drell--Yan and vector mesons.}
    \label{fig:DY_Mll}
\end{figure}

The dielectron invariant mass distribution from Drell--Yan, $~J/\psi\mathrm{,~and~}\Upsilon$ production is shown in Fig.~\ref{fig:DY_Mll}. The dielectron channel is chosen since electrons will likely have a higher purity than muons, at the price of larger radiative corrections. The inelastic $J/\psi$ cross section was measured by H1~\cite{H1:2010udv} and ZEUS~\cite{ZEUS:1997wrc,ZEUS:2002src} to be approximately 2 nanobarns for the range of $W$ relevant to tagged photoproduction at the EIC. The $\psi'$ cross section was measured by ZEUS to be a factor of 5 or so smaller than the $J/\psi$ cross section. The yield of $\Upsilon$ decaying to the dielectron channel is roughly approximated from the elastic HERA data of Refs.~\cite{ZEUS:2009Upsilon}, assuming the elastic and inelastic cross sections are similar, as is the case for $J/\psi$. The cross sections for the $\Upsilon(2\mathrm{S})$ and $\Upsilon(3\mathrm{S})$ are assumed to be 10\% and 5\% of the $\Upsilon(1\mathrm{S})$, respectively. The shape of the $M_{\ell\ell}$ spectrum in Fig.~\ref{fig:DY_Mll} includes the expected resolutions of ePIC. The binning is chosen such that at all points the bin width is at least two times the resolution on $M_{\ell\ell}$. Combinatorial and misidentification backgrounds are excluded, since the degree to which they can be subtracted will depend on the final detector performance.

The expected reach in the plane of $x_\gamma$ and $Q^2=M_{\ell\ell}^2$ for 100 fb$^{-1}$ of 10x275 GeV running is shown in Fig.~\ref{fig:xQ2}. The region of large-$Q^2$ and low-$x_\gamma$ is excluded by the low statistics in 100 fb$^{-1}$, not a kinematic cut. A higher luminosity would therefore somewhat extend the reach. The reach overlaps with the existing $F_2^\gamma$ data from PETRA~\cite{TASSO:1986F2gamma,PLUTO:1981F2gamma,JADE1984_F2gamma}, PEP~\cite{TPC2gamma:1987F2gamma,TPC2G_1987_PRL_F2gamma}, TRISTAN~\cite{AMY:1990F2gamma,AMY:1995F2gammaHighQ2,AMY:1997F2gamma,TOPAZ:1994F2gamma}, and the higher-$x_\gamma$ data from LEP~\cite{L3:1999F2gammaQ2,L3:2005F2gamma,ALEPH:2003F2gamma,OPAL:2002F2gammaLEP2,DELPHI:1996F2gamma}. The shaded regions for the $e^+e^-$ data are approximate. Realistically, the regions of $M_{\ell\ell}\approx M_{J/\psi}$ and $M_{\ell\ell}\approx M_{\Upsilon}$ will likely not be accessible for Drell--Yan.
\begin{figure}[h]
    \centering
    \qquad 
    \includegraphics[width=1.0\linewidth,
        trim={0.1cm 0.1cm 0.1cm 0.1cm},
        clip
    ]{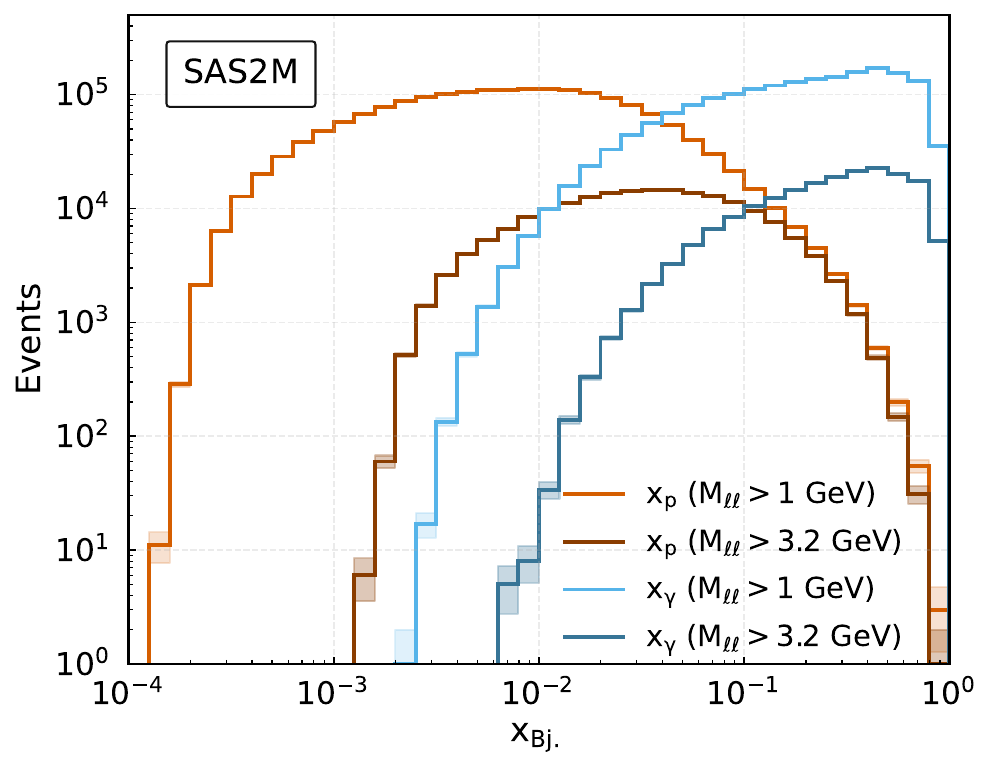}
    \caption{Expected Bjorken $x$ distributions for the proton and photon in tagged Drell--Yan events at the EIC. The darker--colored curves show the effect of cutting on $M_{\ell\ell}>M_{J/\psi}$.}
    \label{fig:DY_xbj}
\end{figure}
\begin{figure}[ht]
    \centering
    \begin{subfigure}[t]{0.95\linewidth}
        \centering
        \includegraphics[width=\linewidth,
            trim={0.1cm 0.1cm 0.1cm 0.1cm},
            clip
        ]{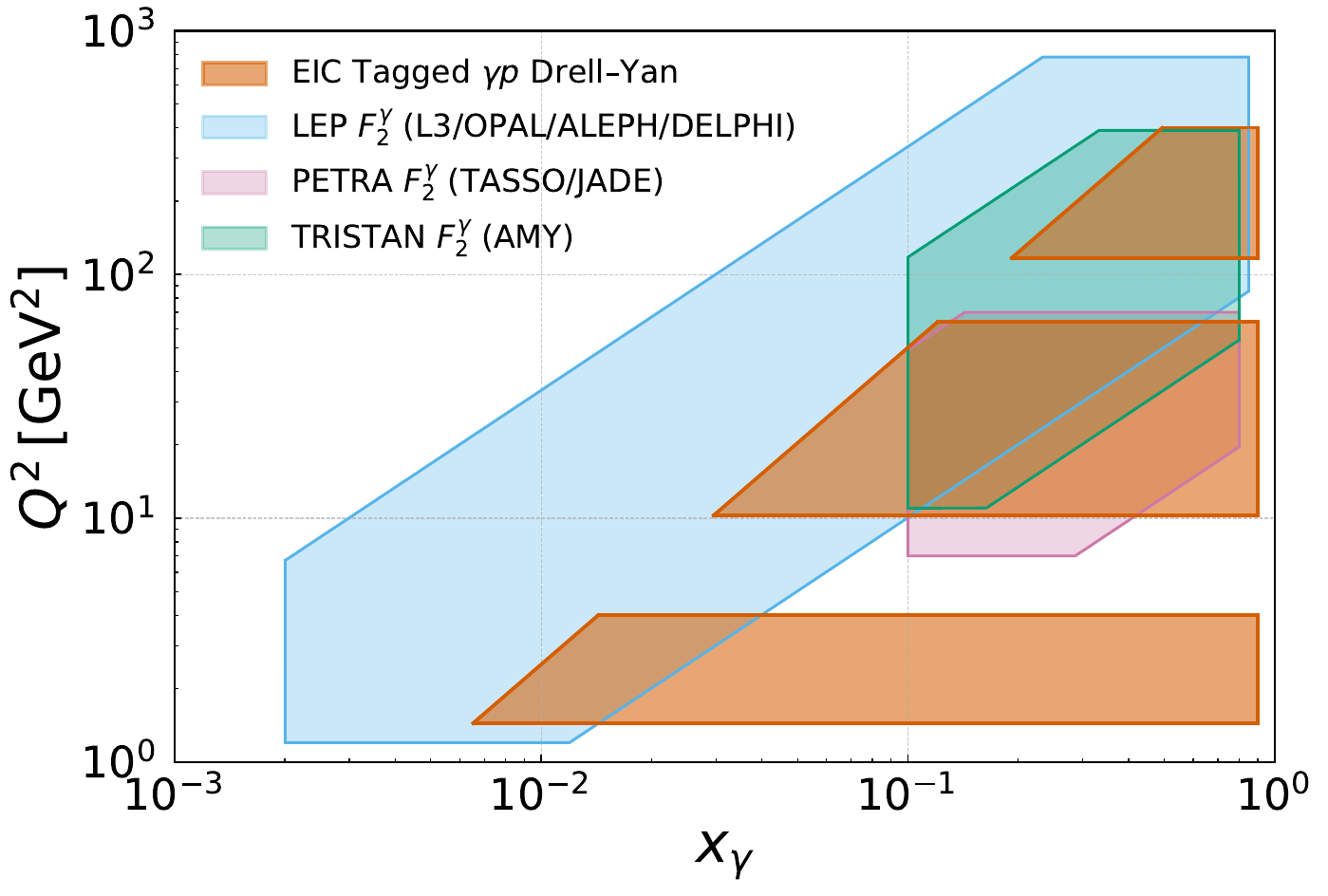}
    \end{subfigure}
    \caption{Projected kinematic coverage in $x_\gamma$ and the hard scale $Q^2=M_{\ell\ell}^2$ of the EIC compared to existing measurements of $F_2^\gamma$ at $e^+e^-$ colliders.}
    \label{fig:xQ2}
\end{figure}

\subsection{Spin Asymmetries}
With the reasonably high expected statistics, the transverse and longitudinal target spin asymmetries of $\gamma\vec{p}\rightarrow\ell^+\ell^-+X$ can also be measured multi-differentially. 

In the end, the precision of the asymmetry measurements will be influenced strongly by the understanding of the asymmetry in the combinatorial and vector meson background. To roughly estimate the uncertainty from backgrounds, we assume an absolute 5\% uncertainty on the target-spin asymmetry of the background, and a signal-to-background ratio that grows logarithmically from 0.1 at $M_{\ell\ell}=1~\mathrm{GeV}$ to 10 at $M_{\ell\ell}=15~\mathrm{GeV}$. No dependence of the background on $p_{T,\ell\ell}$ is assumed. This signal-to-background is roughly consistent with the situation from PHENIX in $p+p$ collisions~\cite{PHENIX:2009gyd,PHENIX:2018dwt}, assuming an extra factor of 5 suppression is achieved at the EIC for semileptonic decays of heavy-flavor hadrons, which form the dominant background. The asymmetry of the background from heavy-flavor can be measured directly and subtracted from the overall yield. Uncorrelated combinatorial background can be dealt with by measuring the rate of like-sign dilepton pairs and their asymmetry.

\begin{figure}[h!]
    \centering
    \begin{subfigure}[t]{0.95\linewidth}
        \centering
        \includegraphics[width=\linewidth,
            trim={0.1cm 0.1cm 0.1cm 0.1cm},
            clip
        ]{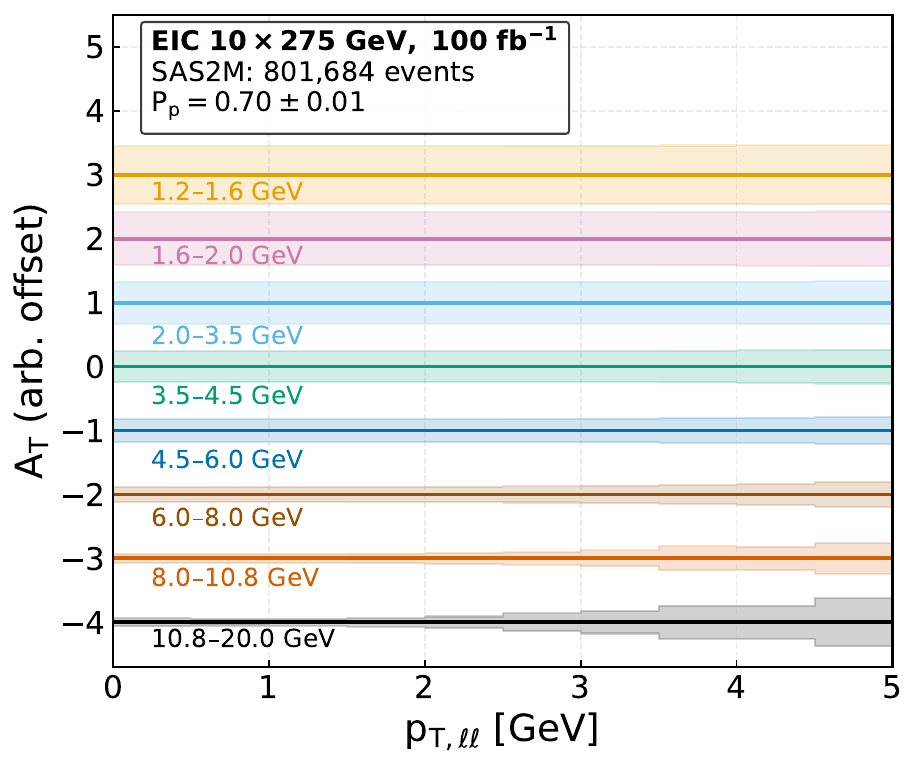}
    \end{subfigure}
    \caption{Estimated uncertainties on the target-spin asymmetries $A_T$ for Drell--Yan at the EIC. Uncertainties are in absolute scale. The different mass bins are offset by one unit on the $y$-axis. The bin widths in $p_{T,\ell\ell}$ are all 0.5 GeV.}
    \label{fig:DYAsym}
\end{figure}

The projections for the measurement of the target-spin asymmetries as a function of pair $p_T$ for several bins of $M_{\ell\ell}$ are shown in Fig.~\ref{fig:DYAsym}. The proton beam polarization is taken to be $70\pm1\%$. The total uncertainty in Fig.~\ref{fig:DYAsym} includes the polarization uncertainty, the uncertainty on the asymmetry of the background, and the statistical uncertainty. The results remain the same whether the target is polarized longitudinally or transversely, assuming the same level of uncertainty can be reached on the asymmetries of the backgrounds. Fig.~\ref{fig:DYAsym} shows that the uncertainty on the asymmetry at low $M_{\ell\ell}$ is large and dominated by the uncertainty from the background while the large $M_{\ell\ell}$ data is reasonably precise since the amount of background is expected to be lower. It should be noted that the $y$-axis scale in Fig.~\ref{fig:DYAsym} is very large: the uncertainty on the most precise asymmetry points, i.e., those at large $M_{\ell\ell}$ and low $p_{T,\ell\ell}$, is around 5\%.

\subsection{Photon TMDPDFs}

In contrast to the proton or pion, studies of the multi-dimensional partonic structure of the photon are still relatively scarce\footnote{The well-informed ArXiv reader should notify the author of any omissions in this regard.}. Most model calculations so far investigate the transverse-momentum-dependent structure of the photon in QED, where the photon fluctuates into lepton pairs and no QCD dynamics is included~\cite{Bacchetta:2016electronPhotonTMD,Hu:2021ElectronBLFQTMD,Kumar:2018ElectronWigner,Nair:2022evk,Nair:2023lir}. The anomalous generalized parton distributions (GPDs) of the photon were shown to be accessible via DVCS on a photon target in Ref.~\cite{Friot:2006mm}. Using overlaps of photon light-front wave functions, Refs.~\cite{Mukherjee:2011bn,Mukherjee:2012zeta,Mukherjee:2013yf,Kumar:2019PhotonLCQM,Gabdrakhmanov:2012PhotonGPD} analyze the GPDs and impact-parameter–dependent distributions of the photon including quark--antiquark degrees of freedom. More recently, light-front quark models have been used to compute the leading-twist TMDs of the photon when it is treated as a $q\bar q$ state~\cite{Puhan:2024qmo}.

The photon provides an especially attractive system in which to test the TMD formalism. In contrast to hadronic targets, whose partonic structure is generated by intrinsically nonperturbative bound-state dynamics and is not calculable from first principles, the source of the photon’s parton densities arises from the perturbatively calculable splitting $\gamma \to q\bar q$ and subsequent QCD radiation. The photon thus offers a relatively well-controlled environment in which the structure and universality of TMD factorization and evolution can be scrutinized. In addition, the TMDPDFs extracted from data can be compared directly to first-principles calculations from the light-front wave functions discussed earlier~\cite{Mukherjee:2011bn,Mukherjee:2012zeta,Mukherjee:2013yf,Kumar:2019PhotonLCQM,Gabdrakhmanov:2012PhotonGPD}. 

For extraction of TMDPDFs, Drell--Yan has the benefit that no fragmentation functions are required, unlike semi-inclusive DIS. However, the Drell--Yan measurement only accesses the convolution of the TMDPDFs of the two incoming beams. Some knowledge of the proton TMDPDFs is thus necessary to disentangle the photon TMDPDFs in Drell--Yan. Thankfully, SIDIS and jet data from the EIC will significantly shrink the uncertainty on the proton TMDPDFs, meaning the photon TMDs can more easily be de-convoluted. A full quantitative projection of the precision with which photon TMDPDFs could be determined from $\gamma p$ Drell--Yan at the EIC unfortunately lies beyond the scope of this work, since a realistic extraction of TMDPDFs generally requires a global analysis, including a heavy reliance on theoretical inputs. This task is therefore better suited to collaborations such as MAP or JAM. Given the fact that the photon is more amenable to theoretical calculations, it could be the case that a precision on the photon TMDs similar to the present precision on the proton TMDs could be achieved even with just the EIC data alone.

\section{Utilizing Interference}
\label{sec:DYBH}

\subsection{Charge-odd lepton asymmetry from Bethe--Heitler--Drell--Yan interference}

Instead of serving merely as a background to the Drell--Yan signal, the Bethe--Heitler process can in principle be turned into a tool to access novel combinations of photon and proton parton distributions via its interference with partonic amplitudes. An analogous situation occurs in DVCS and TCS, where the interference between Bethe--Heitler and the partonic Compton amplitude provides sensitivity to the real and imaginary parts of the Compton form factors. We focus first on the interference between Bethe--Heitler and the standard Drell--Yan mechanism in unpolarized $\gamma p\to \ell^+\ell^- X$, following the setup of Ref.~\cite{Busenitz:1981hr}.

\begin{figure}[h]
    \centering
    \begin{subfigure}[t]{0.49\linewidth}
        \centering
        \includegraphics[width=\linewidth,
            trim={0.1cm 0.1cm 0.1cm 0.1cm},
            clip
        ]{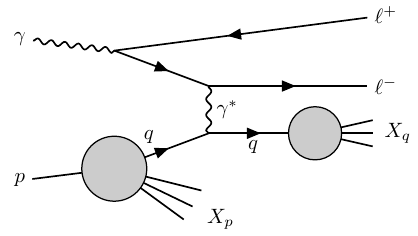}
    \end{subfigure}
    \hfill
    \begin{subfigure}[t]{0.49\linewidth}
        \centering
        \includegraphics[width=\linewidth,
            trim={0.1cm 0.1cm 0.1cm 0.1cm},
            clip
        ]{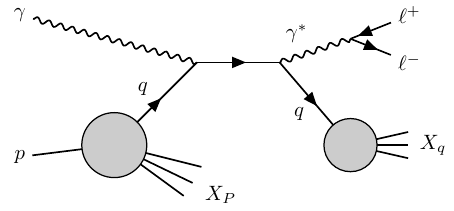}
        \hspace{0.5cm}
    \end{subfigure}
    \begin{subfigure}[t]{0.6\linewidth}
        \centering
        \includegraphics[width=\linewidth,
            trim={0.1cm 0.1cm 0.1cm 0.1cm},
            clip
        ]{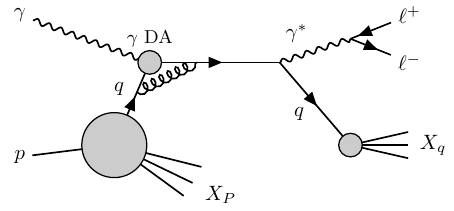}
    \end{subfigure}
    \caption{Diagrams for strongly interfering processes contributing to inclusive dilepton photoproduction plus two jets. \textbf{Top Left:} Bethe--Heitler process. \textbf{Top Right:} Drell--Yan in direct photoproduction ($x_{\gamma}\approx1$). \textbf{Bottom:} Example chiral--odd subprocess involving the photon distribution amplitude.}
    \label{fig:Int}
\end{figure}

The interference between the Bethe--Heitler and Drell--Yan processes is largest in the two–jet topology. For Bethe--Heitler, this corresponds to the process in the top left panel of Fig.~\ref{fig:Int}, where the exchanged photon between the lepton and the proton carries sizable momentum and produces a target jet and a current jet in the final state. In this configuration, the Bethe--Heitler process becomes experimentally indistinguishable, at the level of hadronic activity, from the direct Drell--Yan contribution shown in the top right panel of Fig.~\ref{fig:Int}, in which the incoming photon couples directly to a quark without leaving a spectator remnant in the photon direction. It should be reiterated that this contribution from \emph{direct} photoproduction is not included in the \textsc{Sherpa} simulation discussed in the rest of this paper. However, the ratio of the cross sections for direct and resolved photoproduction of jets at HERA was observed to be of order unity, indicating that the direct yields should be within an order of magnitude of the resolved yields obtained from \textsc{Sherpa}~\cite{ZEUS:2001zoq,H1:2002apm,ZEUS:1995xfc,H1:2006rre}. Resolved photoproduction will also interfere with Bethe--Heitler, but to a lesser degree, since the kinematics of the jets and leptons in resolved photoproduction at $x_{\gamma}\ll1$ are significantly different from those at $x_{\gamma}\approx1$.

In the two–jet configuration, the cross section for dilepton production can be schematically written as
\[
d\sigma \;\sim\; \sum_q \left[ e_q^4\,|A_{\rm DY}|^2 \;+\; e_q^2\,|A_{\rm BH}|^2 \;+\; 2\,e_q^3\,{\rm Re}\left(A_{\rm DY}A_{\rm BH}^\ast\right) \right] ,
\]
where $A_{\rm DY}$ and $A_{\rm BH}$ are the Drell--Yan and Bethe--Heitler amplitudes, respectively, and $e_q$ is the quark electric charge. The pure Drell--Yan piece carries an $e_q^4$ weighting, while the Bethe--Heitler term is proportional to $e_q^2$, as in DIS. The interference term instead scales as $e_q^3$ and is linear in $A_{\rm DY}$. Thus, if it can be isolated experimentally, the interference is sensitive to a different charge-weighted combination of quark distributions in the photon than either inclusive Drell--Yan or DIS. The interference term cancels when integrating over all lepton angles in the dilepton center-of-mass frame, so specific angular observables must be constructed to isolate it.

To provide sensitivity to the Bethe--Heitler–Drell--Yan interference term, one can define an interference–driven angular asymmetry in the dilepton rest frame that is identically zero for the pure Bethe--Heitler and pure Drell--Yan contributions and nonzero only for their interference. We work in the $\ell^+\ell^-$ center-of-mass frame and specify the vector of the positively charged lepton via polar and azimuthal angles $(\theta,\phi)$ with respect to the incoming-photon direction, with the $xz$ plane chosen to contain the proton momentum. Under interchange of the lepton charge one has $(\theta,\phi)\to(\pi-\theta,\phi+\pi)$; the Bethe--Heitler and Drell--Yan squared amplitudes are even under this transformation, whereas the interference term is odd. To define an asymmetry sensitive to the interference, the lepton phase space can be divided into two azimuthal hemispheres, denoted here as left and right, defined by $\phi\in[\pi/2,3\pi/2]$ and $\phi\in[-\pi/2,\pi/2]$, respectively, integrating over the available range in $\cos\theta$. An asymmetry can then be written simply as
\begin{equation}
A_{\rm INT}
= \frac{\sigma_{\rm L}-\sigma_{\rm R}}{\sigma_{\rm L}+\sigma_{\rm R}}\,,
\label{eq:AINT}
\end{equation}
where $\sigma_{\rm L,R}$ is the cross section for having the positive lepton in the left or right hemisphere. By construction, this asymmetry isolates the charge–odd interference term $2\,\mathrm{Re}(A_{\rm DY}A_{\rm BH}^\ast)$. The asymmetry is therefore sensitive to an $e_q^3$–weighted sum over quark flavors, schematically $\sum_q e_q^3\,f_{q/\gamma}(x_\gamma,\mu)\,f_{q/p}(x_p,\mu)\,\mathrm{Re}(A_{\rm DY}$). Measuring $A_{\rm INT}$ as a function of $x_\gamma,x_p,Q^2,p_{T,\ell\ell}$ at the EIC would thus provide access to a unique charge–weighted quark combination that is complementary to those probed in inclusive deep–inelastic scattering and resolved–photon Drell--Yan. The comparison of $e^+e^-$ and $\mu^+\mu^-$ final states offers an additional handle that can be phenomenologically useful, since the purely QED Bethe--Heitler contribution is more sensitive to the lepton mass than the Drell--Yan piece, making a lepton-flavor asymmetry measurement also potentially interesting.

At the EIC, the situation for resolving these interferences is favorable in several respects. Resolved photoproduction Drell--Yan cross sections are at the few-pb level, implying on the order of $10^5$ direct photoproduction dilepton events in $100~\mathrm{fb}^{-1}$. In the resolved photoproduction analysis above, we studied regions in $x_\gamma$, where the Bethe--Heitler contribution is suppressed and the resolved-photon Drell--Yan signal dominates. The interference term also vanishes in that case when integrating over $\theta$ and $\phi$. However, one can deliberately select the complementary kinematic regime where Bethe--Heitler becomes large and the interfering topology is realized, i.e.,\ $x_\gamma\to 1$ with two hard jets. In this region, the high luminosity and large acceptance at the EIC make it feasible to measure charge-odd observables such as $A_{\rm INT}$. The limitation on such an asymmetry measurement will likely be the understanding of the background asymmetry that must be subtracted, similar to the target-spin asymmetry case discussed previously.

\subsection{Access to transversity}

In addition to the unpolarized charge asymmetry $A_{\rm INT}$, transverse target-spin asymmetries have been predicted to be sensitive to the nucleon’s transversity distribution and to the chiral–odd distribution amplitude of the real photon~\cite{Pire:2009ap}. Extractions of the quark transversity distribution in semi-inclusive DIS require coupling the transversity PDF $h_1^q(x)$ to a chiral-odd fragmentation function. In the single-hadron channel this role is played by the Collins fragmentation function $H_1^{\perp,q}(z,p_T^2)$, taken from global fits constrained by SIDIS and $e^+e^-$ data from HERMES, COMPASS, and Belle~\cite{Airapetian:2010ds,Adolph:2012sn,Seidl:2008xc,Anselmino:2007fs,Anselmino:2013vqa,DAlesio:2020vtw}. An alternative method uses dihadron SIDIS and the chiral-odd interference fragmentation function $H_1^{\sphericalangle,q}(z,M_h^2)$. At the EIC, where the transversity PDF is a key physics deliverable, it is desirable to have a way to cross check whether the measured SIDIS asymmetries are a result of the fragmentation functions or the PDFs. Ref.~\cite{Pire:2009ap} proposes the dilepton photoproduction process as a novel avenue for doing precisely this.

Pire and Szymanowski consider the dilepton photoproduction process with a transversely polarized target. Besides the usual Bethe--Heitler and Drell--Yan amplitudes, they introduce a new chiral–odd amplitude $A_\phi$ in which the real photon couples to quarks via its twist–2 distribution amplitude $\phi_\gamma(z)$, normalized to the magnetic susceptibility of the QCD vacuum $\chi\langle q\bar{q}\rangle$. Relevant background on the photon distribution amplitude is provided by Ref.~\cite{Ball:2002ps}. An example process contributing to this amplitude is shown in the bottom of Fig.~\ref{fig:Int}. In this framework the leading-twist transversity-dependent cross section difference,
\begin{equation}
\Delta\sigma_T\ \equiv \sigma^{\uparrow} - \sigma^{\downarrow},
\end{equation}
receives contributions only from the interference of $A_\phi$ with the Bethe--Heitler and standard Drell--Yan amplitudes; gluon radiation does not generate transversity at twist~2 in the massless-quark limit. In the framework of Ref.~\cite{Pire:2009ap},
\begin{equation}
\Delta\sigma_T\ \propto \chi\langle q\bar{q}\rangle \phi_\gamma h^q_1.
\end{equation}
The arguments of $\phi_\gamma(z)$ and $h^q_1$ are suppressed for clarity; the reader should refer to Ref.~\cite{Pire:2009ap} for the details. The magnetic susceptibility and photon distribution amplitude can be extracted from, e.g., QCD sum rules or lattice simulations and used as inputs to relate the asymmetry directly to $h^q_1$. Alternatively, one can take the transversity from SIDIS as an input and attempt to extract the photon distribution amplitude or QCD magnetic susceptibility. This cross section difference can also be turned into an asymmetry for easier experimental measurement.

The highly polarized proton and electron beams at the EIC compel the pursuit of this transversity program in detail. The particle identification capability of the EIC also opens the door for identifying the flavor of the struck quark, providing some access to the flavor-dependent transversity. A full experimental study of the precision on the transversity PDF enabled by this process at the EIC is beyond the scope of this work and will rely on an actual numerical estimate for the cross section difference $\Delta\sigma_T$ in the EIC phase space. Nevertheless, the EIC provides all the tools necessary to make such measurements of transversity, provided the numerical size of the cross section difference is not vanishingly small. While the extraction of transversity will rely on additional non-perturbative inputs, i.e., the photon distribution amplitude and QCD magnetic susceptibility, the lack of dependence on the fragmentation functions breaks the degeneracy that exists in SIDIS, thereby permitting more rigorous extraction of the transversity PDFs from the EIC data if this contribution can be measured.

\section{Pion-photon Drell--Yan}
\begin{figure}[h!]
    \centering
    \begin{subfigure}[t]{0.95\linewidth}
        \centering
        \includegraphics[width=\linewidth,
            trim={0.1cm 0.1cm 0.1cm 0.1cm},
            clip
        ]{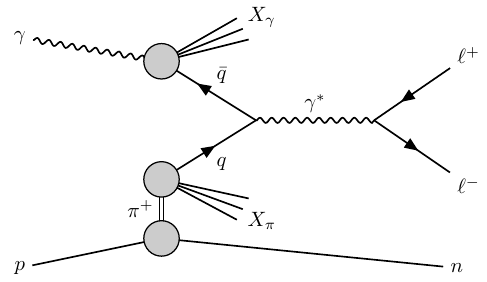}
    \end{subfigure}
    \caption{Diagram of leading order Drell--Yan in $\gamma\pi^+$. This process can be accessed at the EIC by tagging a leading neutron and requiring a large rapidity gap between the neutron and the pion-remnant system $X_\pi$. }
    \label{fig:DYPion}
\end{figure}

Another possibility offered by the EIC is to study Drell--Yan on the pion in leading-neutron photoproduction, i.e., the Sullivan process~\cite{Sullivan:1971kd}\footnote{A similar measurement was recently discussed in the context of ultra-peripheral collisions at the LHC~\cite{Goncalves:2025aay}, although that study focuses on accessing the pion gluon distribution via heavy quark photoproduction.}. In events where a neutron carries a large fraction of the incoming proton beam energy $x_L=E_n/E_p\gtrsim0.8$ and has a small transverse momentum, the HERA data support the so-called proton vertex factorization, whereby the cross section factorizes into a pion flux factor and the hard scattering on a parton inside the pion. Therefore, in the kinematic regime where proton vertex factorization applies, the photoproduction of dileptons and a leading neutron can be interpreted as Drell--Yan between the photon and a $\pi^+$~\cite{Sullivan:1971kd,Kopeliovich:1996iw}. The leading-order diagram of this process is shown in Fig.~\ref{fig:DYPion}. 

Leading-neutron DIS data from HERA, i.e., $ep\rightarrow e'n+X$, provides the core dataset for extractions of the parton distributions of the pion~\cite{ZEUS:1996uej,ZEUS:2002gig,H1:2010hym,Kopeliovich:2012fd,ZEUS:2007knd,McKenney:2015xis,Barry:2018ort,Novikov:2020snp,Cao:2021aci,Barry:2025wjx}. At HERA high-energy leading neutrons were observed in 1-2\% of inclusive DIS events~\cite{ZEUS:2002gig} and 5\% of all dijet photoproduction events~\cite{ZEUS:2000qdw}; a similar percent-level fraction can be expected at the EIC. The one-pion-exchange model describes the ZEUS data well when the leading neutron carries 60\% or more of the incoming proton beam energy~\cite{ZEUS:2002gig}. Therefore, for pions carrying less than 40\% of the incoming proton beam energy, proton vertex factorization seems to hold. JAM imposes a stricter requirement of $x_L>0.8$, corresponding to pions with less than 20\% of the incoming proton beam energy. Applying this requirement, the accessible pion energies for the 10x275 GeV configuration are only 55 GeV and below. It should be remembered that only the region of $1\lesssim E_\gamma \lesssim6~\mathrm{GeV}$ resides within the Low-$Q^2$ tagger acceptance, so the range of center-of-mass energies is only roughly $7\lesssim\sqrt{s_{\gamma\pi}}\lesssim36\mathrm{~GeV}$. This range of energies is similar to existing and proposed fixed-target Drell--Yan experiments at Fermilab, CERN, and JPARC.

The excellent energy and angular resolution of the ePIC Zero-Degree Calorimeter (ZDC) will provide good control over the kinematics of the pion via precise measurements of the leading neutron~\cite{Milton:2024bqv}. The resolution on $x_L$ will be around 2-3\% for the range of $x_L$ of interest. Assuming the fraction of photoproduction Drell--Yan events with leading neutrons is similar to the fraction that was observed in photoproduction and DIS at HERA, one would expect on the order of 10,000 leading-neutron Drell--Yan events in 100 fb$^{-1}$, which should be sufficient for a first measurement. These experimental statistics are not dissimilar to the $\pi^+$ Drell--Yan data on tungsten targets from Refs.~\cite{E615:1989bda,NA10:1986fgk} that have been used in the existing pion PDF fits, although multi-dimensional binning may be limited.

\begin{figure}[h!]
    \centering
    \begin{subfigure}[t]{0.95\linewidth}
        \centering
        \includegraphics[width=\linewidth,
            trim={0.1cm 0.1cm 0.1cm 0.1cm},
            clip
        ]{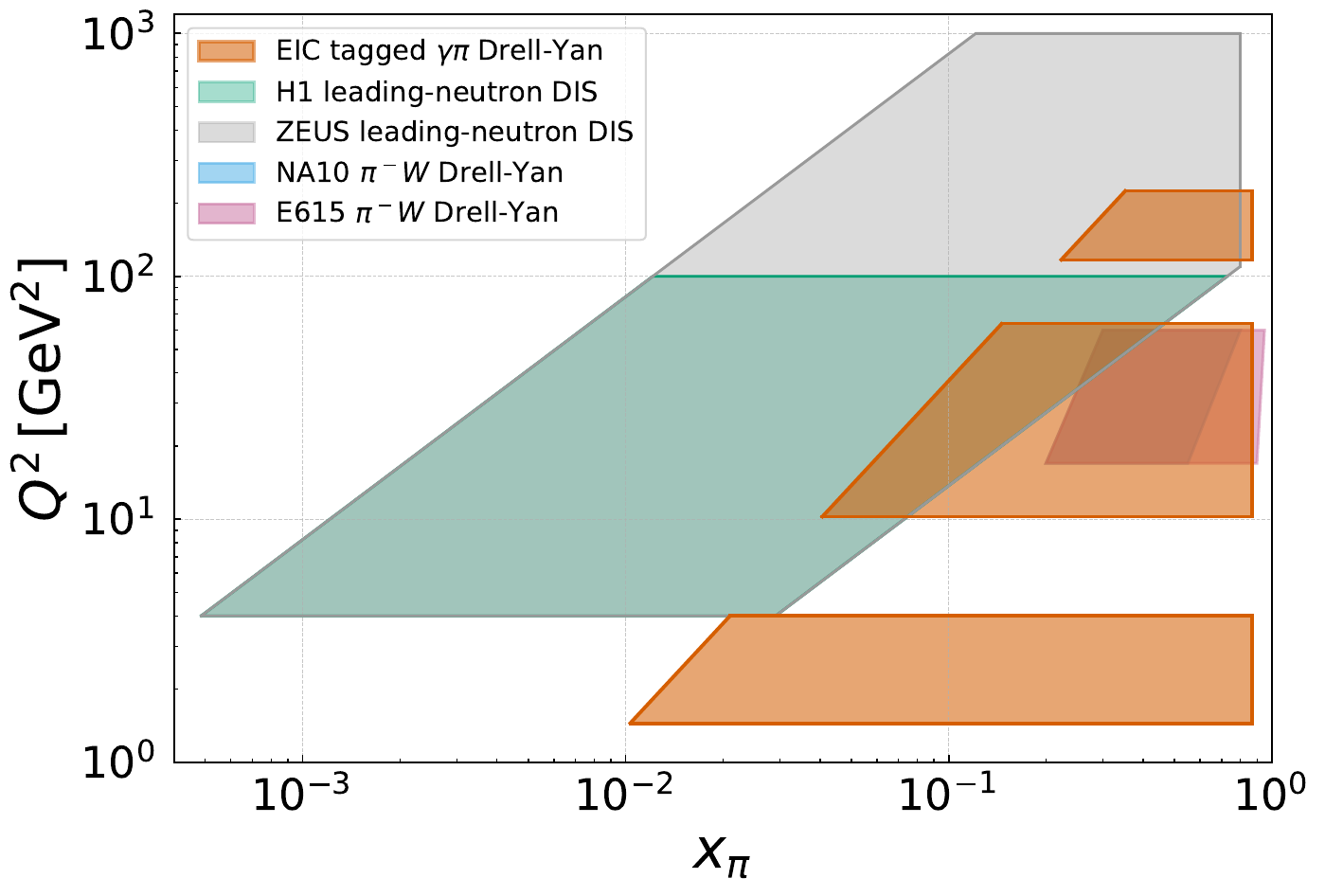}
    \end{subfigure}
    \caption{Projected kinematic coverage of $\gamma\pi^+$ Drell--Yan at the EIC. }
    \label{fig:xQ2_Pi}
\end{figure}

A rough estimate of the accessible kinematic phase space at the EIC compared to existing measurements used in fits of the pion PDFs is shown in Fig.~\ref{fig:xQ2_Pi}. The kinematics can be calculated as $Q^2=sx_{\pi}x_{\gamma}$, where $x_\pi$ is the fraction of the pion's longitudinal momentum that participates in the hard scattering. Kinematics limit the reach to low $x_\pi$, since the center-of-mass energy $s$ is not especially high. At high $Q^2$, it is unlikely that statistics will allow for measurements at $M_{\ell\ell}>M_{\Upsilon}$, although such events do lie within the acceptance.

More speculatively, if leading $\Lambda$ baryons can be reconstructed well in the far-forward region, one could imagine studying the $\gamma K^+$ Drell--Yan process. The feasibility of detecting $\Lambda$ via the neutral decay channel $\Lambda\rightarrow n \pi^0$ was investigated in detail in Ref.~\cite{Paul:2024fww}. The acceptance of the ZDC improves as the energy of the $\Lambda$ increases, so the accessible phase space will correspond to relatively low momentum kaons in the lab frame.

The primary benefits of the $\gamma\pi$ Drell--Yan measurement at the EIC are to test key theoretical assumptions and to access previously unexplored regions of phase space in $Q^2$ and $x_\pi$. It is unlikely that the measurement precision at the EIC will compete with the precision of the leading-neutron DIS or pion Drell--Yan data. However, the universality of proton vertex factorization at collider energies can be tested for the first time in a process other than DIS, helping to validate the assumption that pion structure is accessible in leading-neutron processes more generally.

\section{Conclusion}
\label{sec:conclusion}

The EIC will revolutionize our knowledge of the nucleon, but it also provides an unprecedented opportunity to study the internal structure of the photon via the Drell--Yan process. In photoproduction, Drell--Yan offers a versatile channel to access the partonic content of the photon, complementary to existing constraints from $\gamma\gamma$ and dijet photoproduction measurements.

Using the resolved photoproduction framework implemented in \textsc{Sherpa}, the Drell--Yan cross section at the EIC is predicted to be of order a few picobarns for the $10\times 275$\,GeV configuration, leading to samples of $\mathcal{O}(10^5$--$10^6)$ dilepton pairs in tagged photoproduction for 100 fb$^{-1}$. Measuring this process will require excellent lepton identification, ideally also of muons, across the whole central detector acceptance of $|\eta|<3.5$. If that requirement is met by ePIC or a second EIC detector, the EIC data will enable multi-differential measurements in $M_{\ell\ell}$, $y_{\ell\ell}$, and $p_{T,\ell\ell}$ that map a region in $(x_\gamma,Q^2)$ overlapping and extending beyond existing $F_2^\gamma$ data. 

The $\gamma p$ Drell-Yan data set of the EIC also has the potential to provide the first constraints on transverse-momentum–dependent parton distributions of the photon. Once proton TMDs are better-constrained by SIDIS and jet measurements at the EIC, the $\gamma p$ Drell--Yan process will make it possible to disentangle the photon TMDs. These data can be used to evaluate ab initio theoretical calculations for the partonic structure of the photon from techniques such as the light-front wave function approach. Exploiting the interference between Drell–Yan and Bethe–Heitler opens a handle on novel flavor-weighted quark combinations in collinear PDFs. Furthermore, the chiral-odd amplitude involving the photon distribution amplitude potentially permits access to the nucleon transversity distribution without relying on fragmentation functions. Finally, pion--photon Drell--Yan, accessed via leading-neutron photoproduction, can cross-check and extend the kinematic reach of pion PDF studies. Taken together, these programs demonstrate that Drell--Yan in photoproduction is a promising experimental channel to improve our understanding of QCD with the EIC.

\section{Acknowledgments}
I am grateful to the \textsc{Sherpa} team for implementing resolved photoproduction in their event generator and to P. Meinzinger in particular for help understanding and running \textsc{Sherpa}. Without their efforts this work would not have been possible.

I acknowledge P. Barry, D. Geesaman, M. Grosse Perdekamp, S. Joosten, B. Pire, P. Reimer, L. Szymanowski, and M. Żurek for useful discussions and feedback. I also thank S. Gardner and G. Penman for supplying the Low-$Q^2$ tagger acceptance map. Thanks finally to L. DeWitt for editing this manuscript. This work was supported by the U.S. Department of Energy under Contract No. DE-AC02-06CH11357.

\appendix
\section{Detector Resolutions and Acceptances}\label{sec:appendix}
The strawman detector used for the reconstruction studies follows closely the design of ePIC. When handling charged particles, whichever detector resolution is best is the assumed resolution; i.e., for pions in the barrel, the tracking resolution is used instead of the hadronic calorimetry resolution. $K^0_s$ and $\Lambda$ are assumed to have the momentum resolution of the tracking. Neutrons and $K^0_L$ are produced in roughly equal numbers, so long-lived neutral hadrons are assigned the average of the two masses, i.e., 0.72 GeV. Particle identification of charged hadrons is assumed to be correct within the central detector acceptance. 
\begin{table*}[t]
\centering
\normalsize
\caption{Tracking momentum resolution and acceptance, applied to all charged particles. Global cut: $|\eta|\le 3.5$. Model: $\sigma_p/p=\sqrt{(a\,p)^2+b^2}$ with $p$ in GeV.}
{%
\setlength{\tabcolsep}{12pt}%
\renewcommand{\arraystretch}{1.5}%
\begin{tabular}{|c|c|c|c|c|}
\hline
\textbf{Region ($\eta$)} & \textbf{$a$} & \textbf{$b$} & \textbf{$p_T$ Threshold} & \textbf{Tracking Efficiency} \\
\hline
$-3.5\le\eta<-2.5$  & 0.0010 & 0.030 & $p_T>200$ MeV & 95\%\\
$-2.5\le\eta<-1.0$  & 0.0005 & 0.010 & $p_T>200$ MeV & 95\%\\
$-1.0\le\eta<1.0$   & 0.0005 & 0.005 & $p_T>200$ MeV & 95\%\\
$1.0\le\eta<2.5$    & 0.0005 & 0.010 & $p_T>200$ MeV & 95\%\\
$2.5\le\eta\le 3.5$ & 0.0010 & 0.025 & $p_T>200$ MeV & 95\%\\
\hline
\end{tabular}%
}
\end{table*}

\begin{table*}[t]
\centering
\normalsize
\caption{ECAL energy resolution and threshold, applied to photons. Global cut: $|\eta|\le 3.5$. Model: $\sigma_E/E=\sqrt{S^2/E+C^2}$ with $E$ in GeV.}
{%
\setlength{\tabcolsep}{12pt}%
\renewcommand{\arraystretch}{1.5}%
\begin{tabular}{|l|c|c|l|}
\hline
\textbf{Region ($\eta$)} & \textbf{$S$} & \textbf{$C$} & \textbf{Threshold} \\
\hline
Backward: $-3.5\le\eta<-1.7$ & 0.019 & 0.012 & $E>100$ MeV \\
Central: $-1.7\le\eta\le 1.4$ & 0.049 & 0.001 & $E>100$ MeV \\
Forward: $1.4<\eta\le 3.5$ & 0.106 & 0.032 & $E>100$ MeV \\
\hline
\end{tabular}%
}
\end{table*}

\begin{table*}[t]
\centering
\normalsize
\caption{HCAL energy resolution and threshold, applied to neutral hadrons. Global cut: $|\eta|\le 3.5$. Model: $\sigma_E/E=\sqrt{S^2/E+C^2}$ with $E$ in MeV.}
{%
\setlength{\tabcolsep}{12pt}%
\renewcommand{\arraystretch}{1.5}%
\begin{tabular}{|l|c|c|l|}
\hline
\textbf{Region ($\eta$)} & \textbf{$S$} & \textbf{$C$} & \textbf{Threshold} \\
\hline
Backward: $-3.5\le\eta<-1.4$                & 0.75 & 0.145 & $E>500$ MeV \\
Central: $-1.4\le\eta\le 1.4$        & 0.75 & 0.145 & $E>500$ MeV \\
Forward: $1.4<\eta\le 3.5$  & 0.27 & 0.03  & $E>500$ MeV \\
\hline
\end{tabular}%
}
\end{table*}

\begin{table*}[t]
\centering
\normalsize
\caption{Low-$Q^2$ tagger smearing and acceptance, applied to the scattered electron in the tagged sample.}
{%
\setlength{\tabcolsep}{12pt}%
\renewcommand{\arraystretch}{1.5}%
\begin{tabular}{|l|l|l|}
\hline
\textbf{Quantity} & \textbf{Value} & \textbf{Notes} \\
\hline
Energy resolution   & $\sigma_E/E = 0.0029$        & Gaussian smear \\
Polar angle         & $\sigma_\theta = 0.18$ mrad & Gaussian smear \\
Azimuthal angle     & $\sigma_\phi = 5.4^\circ$    & Gaussian smear \\
$E_e$ acceptance      & See Fig.~\ref{fig:LowQAcceptance}    &  Corresponds to $y\lesssim0.6$\\
\hline
\end{tabular}%
}
\end{table*}

\clearpage
\bibliography{ref} 

@article{TPC2G_1987_PRL_F2gamma,
  author       = {H. Aihara and et al.},
  collaboration= {TPC/Two-Gamma},
  title        = {Measurement of the Photon Structure Function $F_2^{\gamma}$},
  journal      = {Phys. Rev. Lett.},
  volume       = {58},
  pages        = {97--100},
  year         = {1987},
  doi          = {10.1103/PhysRevLett.58.97}
}

@article{ZEUS:2000qdw,
    author = "Breitweg, J. and others",
    collaboration = "ZEUS",
    title = "{Measurement of dijet cross-sections for events with a leading neutron in photoproduction at HERA}",
    eprint = "hep-ex/0010019",
    archivePrefix = "arXiv",
    reportNumber = "DESY-00-142",
    doi = "10.1016/S0550-3213(00)00612-X",
    journal = "Nucl. Phys. B",
    volume = "596",
    pages = "3--29",
    year = "2001"
}

@article{DAlesio:2020vtw,
    author = "D'Alesio, Umberto and Flore, Carlo and Prokudin, Alexei",
    title = "{Role of the Soffer bound in determination of transversity and the tensor charge}",
    eprint = "2001.01573",
    archivePrefix = "arXiv",
    primaryClass = "hep-ph",
    reportNumber = "JLAB-THY-19-3130",
    doi = "10.1016/j.physletb.2020.135347",
    journal = "Phys. Lett. B",
    volume = "803",
    pages = "135347",
    year = "2020"
}

@article{Bawa:1993qr,
    author = "Bawa, A. C. and Charchula, K. and Stirling, W. James",
    title = "{Photoproduction of large mass lepton pairs at HERA as a probe of the small x structure of the proton}",
    eprint = "hep-ph/9307340",
    archivePrefix = "arXiv",
    reportNumber = "DTP-93-44, UDWPHYS-93-02",
    doi = "10.1016/0370-2693(93)90020-I",
    journal = "Phys. Lett. B",
    volume = "313",
    pages = "461--468",
    year = "1993"
}

@article{JADE1984_F2gamma,
  author       = {W. Bartel and et al.},
  collaboration= {JADE},
  title        = {Experimental study of the photon structure function $F_2^{\gamma}$ at high $Q^2$},
  journal      = {Z. Phys. C},
  volume       = {24},
  pages        = {231--247},
  year         = {1984}
}

@article{OPAL1997_ZPC74_AnalysisF2gamma,
  author       = {K. Ackerstaff and et al.},
  collaboration= {OPAL},
  title        = {Analysis of hadronic final states and the photon structure function $F_2^{\gamma}$ in DIS $e\gamma$ scattering at LEP},
  journal      = {Z. Phys. C},
  volume       = {74},
  pages        = {33--48},
  year         = {1997}
}

@article{OPAL1997_Q2Evolution_arXiv,
  author       = {K. Ackerstaff and et al.},
  collaboration= {OPAL},
  title        = {Measurement of the $Q^2$ evolution of the photon structure function $F_2^{\gamma}$},
  journal      = {Phys. Lett. B},
  note         = {arXiv:hep-ex/9708019},
  year         = {1997}
}

@article{OPAL2002_F2cgamma_PLB539,
  author       = {G. Abbiendi and et al.},
  collaboration= {OPAL},
  title        = {Measurement of the charm structure function $F_{2,c}^{\gamma}$ of the photon at LEP},
  journal      = {Phys. Lett. B},
  volume       = {539},
  pages        = {13--24},
  year         = {2002},
  doi          = {10.1016/S0370-2693(02)01999-8}
}

@article{OPAL2000_Dstar_gammaGamma_EPJC16,
  author       = {G. Abbiendi and et al.},
  collaboration= {OPAL},
  title        = {Inclusive production of $D^{*\pm}$ mesons in $\gamma\gamma$ collisions at $\sqrt{s_{ee}}=183$ and $189$ GeV and a first measurement of $F_{2,c}^{\gamma}$},
  journal      = {Eur. Phys. J. C},
  volume       = {16},
  pages        = {579--596},
  year         = {2000},
  doi          = {10.1007/s100520000420}
}

@article{L3_1999_2000_Q2Evolution_F2gamma,
  author       = {M. Acciarri and et al.},
  collaboration= {L3},
  title        = {The $Q^2$ evolution of the hadronic photon structure function $F_2^{\gamma}$},
  journal      = {Phys. Lett. B},
  volume       = {436},
  pages        = {403--414},
  year         = {1998}
}

@article{L3_2004_InclusiveJet_gammagamma_PLB602,
  author       = {P. Achard and et al.},
  collaboration= {L3},
  title        = {Inclusive jet production in two-photon collisions at LEP},
  journal      = {Phys. Lett. B},
  volume       = {602},
  pages        = {157--166},
  year         = {2004}
}

@article{Cornet:2004nb,
    author = "Cornet, F. and Jankowski, P. and Krawczyk, M.",
    title = "{A New 5 flavor NLO analysis and parametrizations of parton distributions of the real photon}",
    eprint = "hep-ph/0404063",
    archivePrefix = "arXiv",
    doi = "10.1103/PhysRevD.70.093004",
    journal = "Phys. Rev. D",
    volume = "70",
    pages = "093004",
    year = "2004"
}

@article{Gluck:1991jc,
    author = "Gluck, M. and Reya, E. and Vogt, A.",
    title = "{Photonic parton distributions}",
    reportNumber = "DO-TH-91-31",
    doi = "10.1103/PhysRevD.46.1973",
    journal = "Phys. Rev. D",
    volume = "46",
    pages = "1973--1979",
    year = "1992"
}

@article{Chu:2017mnm,
    author = "Chu, Xiaoxuan and Aschenauer, Elke-Caroline and Lee, Jeong-Hun and Zheng, Liang",
    title = "{Photon structure studied at an Electron Ion Collider}",
    eprint = "1705.08831",
    archivePrefix = "arXiv",
    primaryClass = "nucl-ex",
    doi = "10.1103/PhysRevD.96.074035",
    journal = "Phys. Rev. D",
    volume = "96",
    number = "7",
    pages = "074035",
    year = "2017"
}

@article{PHENIX:2018dwt,
    author = "Aidala, C. and others",
    collaboration = "PHENIX",
    title = "{Measurements of $\mu\mu$ pairs from open heavy flavor and Drell-Yan in $p+p$ collisions at $\sqrt{s}=200$ GeV}",
    eprint = "1805.02448",
    archivePrefix = "arXiv",
    primaryClass = "hep-ex",
    doi = "10.1103/PhysRevD.99.072003",
    journal = "Phys. Rev. D",
    volume = "99",
    number = "7",
    pages = "072003",
    year = "2019"
}

@article{ZEUS:2002src,
    author = "Chekanov, S. and others",
    collaboration = "ZEUS",
    title = "{Measurements of inelastic J / psi and psi-prime photoproduction at HERA}",
    eprint = "hep-ex/0211011",
    archivePrefix = "arXiv",
    reportNumber = "DESY-02-163",
    doi = "10.1140/epjc/s2002-01130-2",
    journal = "Eur. Phys. J. C",
    volume = "27",
    pages = "173--188",
    year = "2003"
}

@article{Jacobs:2025rqa,
    author = "Jacobs, William W. and Visser, Gerard and Kelleher, Rowan and Vossen, Anselm and Schneider, Simon and Ilieva, Yordanka and Turonski, Pawel",
    title = "{Design and Expected Performance for an hKLM at the EIC}",
    eprint = "2511.08432",
    archivePrefix = "arXiv",
    primaryClass = "physics.ins-det",
    month = "11",
    year = "2025"
}

@article{Paul:2024fww,
    author = "Paul, Sebouh J. and Milton, Ryan and Mor{\'a}n, Sebasti{\'a}n and Schmookler, Barak and Arratia, Miguel",
    title = "{Feasibility study of measuring {\ensuremath{\Lambda}}0{\textrightarrow}n{\ensuremath{\pi}}0 using a high-granularity zero-degree calorimeter at the future electron-ion collider}",
    eprint = "2412.12346",
    archivePrefix = "arXiv",
    primaryClass = "nucl-ex",
    doi = "10.1103/q7w9-sbsc",
    journal = "Phys. Rev. D",
    volume = "111",
    number = "9",
    pages = "092013",
    year = "2025"
}

@article{Friot:2006mm,
    author = "Friot, S. and Pire, B. and Szymanowski, L.",
    title = "{Deeply virtual compton scattering on a photon and generalized parton distributions in the photon}",
    eprint = "hep-ph/0611176",
    archivePrefix = "arXiv",
    reportNumber = "CPHT-RR083-1006, UB-ECM-PF06-32",
    doi = "10.1016/j.physletb.2006.12.038",
    journal = "Phys. Lett. B",
    volume = "645",
    pages = "153--160",
    year = "2007"
}

@article{Pire:2009ap,
    author = "Pire, B. and Szymanowski, L.",
    title = "{Probing the nucleon's transversity and the photon's distribution amplitude in lepton pair photoproduction}",
    eprint = "0905.1258",
    archivePrefix = "arXiv",
    primaryClass = "hep-ph",
    reportNumber = "CPHT-RR034.0409",
    doi = "10.1103/PhysRevLett.103.072002",
    journal = "Phys. Rev. Lett.",
    volume = "103",
    pages = "072002",
    year = "2009"
}

@article{Airapetian:2010ds,
  author         = "Airapetian, A. and others",
  title          = "{Effects of transversity in deep-inelastic scattering by polarized protons}",
  collaboration  = "HERMES",
  journal        = "Phys. Lett. B",
  volume         = "693",
  year           = "2010",
  pages          = "11--16",
  doi            = "10.1016/j.physletb.2010.08.012",
  eprint         = "1006.4221",
  archivePrefix  = "arXiv"
}

@article{Adolph:2012sn,
  author         = "Adolph, C. and others",
  title          = "{Experimental investigation of transverse spin asymmetries in muon-p SIDIS processes: Collins asymmetries}",
  collaboration  = "COMPASS",
  journal        = "Phys. Lett. B",
  volume         = "717",
  year           = "2012",
  pages          = "376--382",
  doi            = "10.1016/j.physletb.2012.09.055",
  eprint         = "1205.5121",
  archivePrefix  = "arXiv"
}

@article{Seidl:2008xc,
  author         = "Seidl, R. and others",
  title          = "{Measurement of azimuthal asymmetries in inclusive production of hadron pairs in e+ e- annihilation at Belle}",
  collaboration  = "Belle",
  journal        = "Phys. Rev. D",
  volume         = "78",
  year           = "2008",
  pages          = "032011",
  doi            = "10.1103/PhysRevD.78.032011",
  eprint         = "0805.2975",
  archivePrefix  = "arXiv"
}

@article{DaRold:2024ram,
    author = "Da Rold, Leandro and Medina, Anibal D. and Roy, Subhojit and Wagner, Carlos E. M.",
    title = "{Testing the lepton content of the proton at HERA and EIC}",
    eprint = "2408.12644",
    archivePrefix = "arXiv",
    primaryClass = "hep-ph",
    reportNumber = "EFI 24-8",
    doi = "10.1007/JHEP01(2025)121",
    journal = "JHEP",
    volume = "01",
    pages = "121",
    year = "2025"
}

@article{Goncalves:2025aay,
    author = "Goncalves, Victor P. and de Souza, Juciene T. and Spiering, Diego",
    title = "{Probing the pion gluon distribution at small - $x$ in photon-induced interactions at LHC}",
    eprint = "2512.13257",
    archivePrefix = "arXiv",
    primaryClass = "hep-ph",
    month = "12",
    year = "2025"
}

@article{Peng:2025etb,
    author = "Peng, Jen-Chieh and Liu, Ming-Xiong and Xu, Guanghua",
    title = "{Sign Reversal of Boer-Mulders Functions from Semi-inclusive Deep-Inelastic Scattering to the Drell-Yan Process}",
    eprint = "2512.12955",
    archivePrefix = "arXiv",
    primaryClass = "hep-ph",
    month = "12",
    year = "2025"
}

@article{Brei:2025isx,
    author = "Brei, Nathan and Jeske, Torri and Lawrence, David",
    title = "{Reconstruction framework advancements to support streaming for the ePIC detector at the EIC}",
    doi = "10.1051/epjconf/202533701189",
    journal = "EPJ Web Conf.",
    volume = "337",
    pages = "01189",
    year = "2025"
}

@article{Ameli:2022gvw,
    author = "Ameli, Fabrizio and others",
    title = "{Streaming readout for next generation electron scattering experiments}",
    eprint = "2202.03085",
    archivePrefix = "arXiv",
    primaryClass = "physics.ins-det",
    reportNumber = "JLAB-PHY-22-3556",
    doi = "10.1140/epjp/s13360-022-03146-z",
    journal = "Eur. Phys. J. Plus",
    volume = "137",
    number = "8",
    pages = "958",
    year = "2022"
}

@article{AbdulKhalek:2021gbh,
    author = "Abdul Khalek, R. and others",
    title = "{Science Requirements and Detector Concepts for the Electron-Ion Collider}: {EIC Yellow Report}",
    eprint = "2103.05419",
    archivePrefix = "arXiv",
    primaryClass = "physics.ins-det",
    reportNumber = "BNL-220990-2021-FORE, JLAB-PHY-21-3198, LA-UR-21-20953",
    doi = "10.1016/j.nuclphysa.2022.122447",
    journal = "Nucl. Phys. A",
    volume = "1026",
    pages = "122447",
    year = "2022"
}

@article{Anselmino:2007fs,
  author         = "Anselmino, M. and Boglione, M. and D'Alesio, U. and Kotzinian, A. and Melis, S. and Murgia, F. and Prokudin, A.",
  title          = "{Transversity and Collins functions from SIDIS and e+ e- data}",
  journal        = "Phys. Rev. D",
  volume         = "75",
  year           = "2007",
  pages          = "054032",
  doi            = "10.1103/PhysRevD.75.054032",
  eprint         = "hep-ph/0701006",
  archivePrefix  = "arXiv"
}

@article{Ball:2002ps,
    author = "Ball, Patricia and Braun, V. M. and Kivel, N.",
    title = "{Photon distribution amplitudes in QCD}",
    eprint = "hep-ph/0207307",
    archivePrefix = "arXiv",
    reportNumber = "IPPP-02-40, DCPT-02-80",
    doi = "10.1016/S0550-3213(02)01017-9",
    journal = "Nucl. Phys. B",
    volume = "649",
    pages = "263--296",
    year = "2003"
}

@article{Anselmino:2013vqa,
  author         = "Anselmino, M. and others",
  title          = "{Simultaneous extraction of transversity and Collins functions from new SIDIS and e+ e- data}",
  journal        = "Phys. Rev. D",
  volume         = "87",
  year           = "2013",
  pages          = "094019",
  doi            = "10.1103/PhysRevD.87.094019",
  eprint         = "1303.3822",
  archivePrefix  = "arXiv"
}

@article{Bacchetta:2016electronPhotonTMD,
  author       = {Bacchetta, Alessandro and Mantovani, Luca and Pasquini, Barbara},
  title        = {The Electron in Three-Dimensional Momentum Space},
  journal      = {Phys. Rev. D},
  volume       = {93},
  pages        = {013005},
  year         = {2016},
  doi          = {10.1103/PhysRevD.93.013005},
  eprint       = {1508.06964},
  archivePrefix= {arXiv},
  primaryClass = {hep-ph}
}

@article{Nair:2023lir,
    author = "Nair, Sreeraj and Mondal, Chandan and Zhao, Xingbo and Mukherjee, Asmita and Vary, James P.",
    collaboration = "BLFQ",
    title = "{Massless and massive photons within basis light-front quantization}",
    eprint = "2302.13645",
    archivePrefix = "arXiv",
    primaryClass = "hep-ph",
    doi = "10.1103/PhysRevD.108.116005",
    journal = "Phys. Rev. D",
    volume = "108",
    number = "11",
    pages = "116005",
    year = "2023"
}

@article{Kumar:2018ElectronWigner,
  author       = {Kumar, Narinder and Mondal, Chandan},
  title        = {Wigner distributions for an electron},
  journal      = {Nucl.\ Phys.\ B},
  volume       = {931},
  pages        = {226--249},
  year         = {2018},
  doi          = {10.1016/j.nuclphysb.2018.04.014},
  eprint       = {1705.03183},
  archivePrefix= {arXiv},
  primaryClass = {hep-ph}
}

@inproceedings{Kumar:2019PhotonLCQM,
  author       = {Kumar, Narinder},
  title        = {Generalized Parton Distributions of the Photon in Light-Cone Quark Model},
  booktitle    = {JPS Conf.\ Proc.},
  volume       = {26},
  pages        = {021007},
  year         = {2019},
  doi          = {10.7566/JPSCP.26.021007}
}

@article{Mukherjee:2012zeta,
  author       = {Mukherjee, Asmita and Nair, Sreeraj},
  title        = {Generalized Parton Distributions of the Photon for Nonzero $\zeta$},
  journal      = {Phys.\ Lett.\ B},
  volume       = {707},
  pages        = {99--106},
  year         = {2012},
  doi          = {10.1016/j.physletb.2011.12.011},
  eprint       = {1110.5242},
  archivePrefix= {arXiv},
  primaryClass = {hep-ph}
}

@article{Hu:2021ElectronBLFQTMD,
  author       = {Hu, Zhi and Xu, Siqi and Mondal, Chandan and Zhao, Xingbo and Vary, James P.},
  title        = {Transverse structure of electron in momentum space in basis light-front quantization},
  journal      = {Phys.\ Rev.\ D},
  volume       = {103},
  number       = {3},
  pages        = {036005},
  year         = {2021},
  doi          = {10.1103/PhysRevD.103.036005},
  eprint       = {2010.12498},
  archivePrefix= {arXiv},
  primaryClass = {hep-ph}
}

@article{Mukherjee:2013yf,
    author = "Mukherjee, Asmita and Nair, Sreeraj and Kumar Ojha, Vikash",
    title = "{Generalized Parton Distributions of the Photon with Helicity Flip}",
    eprint = "1301.6895",
    archivePrefix = "arXiv",
    primaryClass = "hep-ph",
    doi = "10.1016/j.physletb.2013.03.031",
    journal = "Phys. Lett. B",
    volume = "721",
    pages = "284--289",
    year = "2013"
}

@article{Mukherjee:2011bn,
    author = "Mukherjee, Asmita and Nair, Sreeraj",
    title = "{Generalized Parton Distributions of the Photon}",
    eprint = "1105.5299",
    archivePrefix = "arXiv",
    primaryClass = "hep-ph",
    doi = "10.1016/j.physletb.2011.11.004",
    journal = "Phys. Lett. B",
    volume = "706",
    pages = "77--81",
    year = "2011"
}

@article{Puhan:2024qmo,
    author = "Puhan, Satyajit and Kumar, Narinder and Dahiya, Harleen",
    title = "{Photon Leading Twist Transverse Momentum Dependent Parton Distributions}",
    eprint = "2408.07714",
    archivePrefix = "arXiv",
    primaryClass = "hep-ph",
    doi = "10.1140/epja/s10050-025-01527-3",
    journal = "Eur. Phys. J. A",
    volume = "61",
    number = "3",
    pages = "56",
    year = "2025"
}

@article{Nair:2022evk,
    author = "Nair, Sreeraj and Mondal, Chandan and Zhao, Xingbo and Mukherjee, Asmita and Vary, James P.",
    collaboration = "BLFQ",
    title = "{Basis light-front quantization approach to photon}",
    eprint = "2201.12770",
    archivePrefix = "arXiv",
    primaryClass = "hep-ph",
    doi = "10.1016/j.physletb.2022.137005",
    journal = "Phys. Lett. B",
    volume = "827",
    pages = "137005",
    year = "2022"
}

@article{E615:1989bda,
    author = "Conway, J. S. and others",
    collaboration = "E615",
    title = "{Experimental Study of Muon Pairs Produced by 252-GeV Pions on Tungsten}",
    reportNumber = "FERMILAB-PUB-89-0304-V",
    doi = "10.1103/PhysRevD.39.92",
    journal = "Phys. Rev. D",
    volume = "39",
    pages = "92--122",
    year = "1989"
}

@article{Gabdrakhmanov:2012PhotonGPD,
  author       = {Gabdrakhmanov, I.~R. and Teryaev, O.~V.},
  title        = {Analyticity and sum rules for photon GPDs},
  journal      = {Phys.\ Lett.\ B},
  volume       = {716},
  pages        = {417--424},
  year         = {2012},
  doi          = {10.1016/j.physletb.2012.08.041},
  eprint       = {1204.6471},
  archivePrefix= {arXiv},
  primaryClass = {hep-ph}
}

@article{NA10:1986fgk,
    author = "Falciano, S. and others",
    collaboration = "NA10",
    title = "{Angular Distributions of Muon Pairs Produced by 194-{GeV}/$c$ Negative Pions}",
    reportNumber = "CERN-EP-86-35",
    doi = "10.1007/BF01551072",
    journal = "Z. Phys. C",
    volume = "31",
    pages = "513",
    year = "1986"
}

@article{ZEUS:2002gig,
    author = "Chekanov, S. and others",
    collaboration = "ZEUS",
    title = "{Leading neutron production in e+ p collisions at HERA}",
    eprint = "hep-ex/0205076",
    archivePrefix = "arXiv",
    reportNumber = "DESY-02-039",
    doi = "10.1016/S0550-3213(02)00439-X",
    journal = "Nucl. Phys. B",
    volume = "637",
    pages = "3--56",
    year = "2002"
}

@article{McKenney:2015xis,
    author = "McKenney, J. R. and Sato, Nobuo and Melnitchouk, W. and Ji, Chueng-Ryong",
    title = "{Pion structure function from leading neutron electroproduction and SU(2) flavor asymmetry}",
    eprint = "1512.04459",
    archivePrefix = "arXiv",
    primaryClass = "hep-ph",
    reportNumber = "JLAB-THY-15-2183",
    doi = "10.1103/PhysRevD.93.054011",
    journal = "Phys. Rev. D",
    volume = "93",
    number = "5",
    pages = "054011",
    year = "2016"
}

@article{Kopeliovich:1996iw,
    author = "Kopeliovich, Boris and Povh, Bogdan and Potashnikova, Irina",
    title = "{Deep inelastic electroproduction of neutrons in the proton fragmentation region}",
    eprint = "hep-ph/9601291",
    archivePrefix = "arXiv",
    reportNumber = "DESY-96-011",
    doi = "10.1007/s002880050301",
    journal = "Z. Phys. C",
    volume = "73",
    pages = "125--131",
    year = "1996"
}

@article{Milton:2024bqv,
    author = "Milton, Ryan and Paul, Sebouh J. and Schmookler, Barak and Arratia, Miguel and Karande, Piyush and Angerami, Aaron and Acosta, Fernando Torales and Nachman, Benjamin",
    title = "{Design and simulation of a SiPM-on-tile ZDC for the future EIC, and its performance with graph neural networks}",
    eprint = "2406.12877",
    archivePrefix = "arXiv",
    primaryClass = "physics.ins-det",
    doi = "10.1016/j.nima.2025.170613",
    journal = "Nucl. Instrum. Meth. A",
    volume = "1079",
    pages = "170613",
    year = "2025"
}

@article{Kopeliovich:2012fd,
    author = "Kopeliovich, B. Z. and Potashnikova, I. K. and Povh, B. and Schmidt, Ivan",
    title = "{Pion structure function at small x from DIS data}",
    eprint = "1205.0067",
    archivePrefix = "arXiv",
    primaryClass = "hep-ph",
    doi = "10.1103/PhysRevD.85.114025",
    journal = "Phys. Rev. D",
    volume = "85",
    pages = "114025",
    year = "2012"
}

@article{ZEUS:1996uej,
    author = "Derrick, M. and others",
    collaboration = "ZEUS",
    title = "{Observation of events with an energetic forward neutron in deep inelastic scattering at HERA}",
    eprint = "hep-ex/9606006",
    archivePrefix = "arXiv",
    reportNumber = "DESY-96-093",
    doi = "10.1016/0370-2693(96)00688-0",
    journal = "Phys. Lett. B",
    volume = "384",
    pages = "388--400",
    year = "1996"
}

@article{ZEUS:2007knd,
    author = "Chekanov, S. and others",
    collaboration = "ZEUS",
    title = "{Leading neutron energy and pT distributions in deep inelastic scattering and photoproduction at HERA}",
    eprint = "hep-ex/0702028",
    archivePrefix = "arXiv",
    reportNumber = "DESY-07-011",
    doi = "10.1016/j.nuclphysb.2007.03.045",
    journal = "Nucl. Phys. B",
    volume = "776",
    pages = "1--37",
    year = "2007"
}

@article{H1:2010hym,
    author = "Aaron, F. D. and others",
    collaboration = "H1",
    title = "{Measurement of Leading Neutron Production in Deep-Inelastic Scattering at HERA}",
    eprint = "1001.0532",
    archivePrefix = "arXiv",
    primaryClass = "hep-ex",
    reportNumber = "DESY-09-185",
    doi = "10.1140/epjc/s10052-010-1369-4",
    journal = "Eur. Phys. J. C",
    volume = "68",
    pages = "381--399",
    year = "2010"
}

@article{Barry:2018ort,
    author = "Barry, P. C. and Sato, N. and Melnitchouk, W. and Ji, Chueng-Ryong",
    title = "{First Monte Carlo Global QCD Analysis of Pion Parton Distributions}",
    eprint = "1804.01965",
    archivePrefix = "arXiv",
    primaryClass = "hep-ph",
    reportNumber = "JLAB-THY-18-2678",
    doi = "10.1103/PhysRevLett.121.152001",
    journal = "Phys. Rev. Lett.",
    volume = "121",
    number = "15",
    pages = "152001",
    year = "2018"
}

@article{Cao:2021aci,
    author = "Cao, N. Y. and Barry, P. C. and Sato, N. and Melnitchouk, W.",
    collaboration = "Jefferson Lab Angular Momentum",
    title = "{Towards the three-dimensional parton structure of the pion: Integrating transverse momentum data into global QCD analysis}",
    eprint = "2103.02159",
    archivePrefix = "arXiv",
    primaryClass = "hep-ph",
    reportNumber = "JLAB-THY-21-3328",
    doi = "10.1103/PhysRevD.103.114014",
    journal = "Phys. Rev. D",
    volume = "103",
    number = "11",
    pages = "114014",
    year = "2021"
}

@article{Barry:2025wjx,
    author = "Barry, P. C. and Ji, Chueng-Ryong and Melnitchouk, W. and Sato, N. and Steffens, Fernanda",
    collaboration = "JAM",
    title = "{First simultaneous global QCD analysis of kaon and pion parton distributions with lattice QCD constraints}",
    eprint = "2510.11979",
    archivePrefix = "arXiv",
    primaryClass = "hep-ph",
    reportNumber = "JLAB-THY-25-4569",
    month = "10",
    year = "2025"
}

@article{Novikov:2020snp,
    author = "Novikov, Ivan and others",
    title = "{Parton Distribution Functions of the Charged Pion Within The xFitter Framework}",
    eprint = "2002.02902",
    archivePrefix = "arXiv",
    primaryClass = "hep-ph",
    reportNumber = "DESY-20-013, DESY 20-013",
    doi = "10.1103/PhysRevD.102.014040",
    journal = "Phys. Rev. D",
    volume = "102",
    number = "1",
    pages = "014040",
    year = "2020"
}

@article{Deng:2025hio,
    author = "Deng, Yongjie and Jiang, Xu-Hui and Liu, Tianbo and Yan, Bin",
    title = "{Testing lepton flavor universality at the Electron-Ion Collider}",
    eprint = "2503.02605",
    archivePrefix = "arXiv",
    primaryClass = "hep-ph",
    doi = "10.1007/JHEP06(2025)157",
    journal = "JHEP",
    volume = "06",
    pages = "157",
    year = "2025"
}

@article{Zhang:2023EICetau,
  author       = {Zhang, J.-L. and Mantry, S. and Adkins, J. K. and others},
  title        = {Search for e→τ charged lepton flavor violation at the {EIC} with the {ECCE} detector},
  journal      = {Nucl. Instrum. Meth. A},
  volume       = {1053},
  pages        = {168276},
  year         = {2023},
  doi          = {10.1016/j.nima.2023.168276},
  archivePrefix= {arXiv},
  eprint       = {2207.10261},
  primaryClass = {hep-ex}
}

@article{Chwastowski:2022fzk,
    author = "Chwastowski, Janusz J. and Piotrzkowski, Krzysztof and Przybycien, Mariusz",
    title = "{Exclusive lepton pair production at the electron{\textendash}ion collider}",
    eprint = "2206.02466",
    archivePrefix = "arXiv",
    primaryClass = "hep-ph",
    doi = "10.1140/epjc/s10052-022-10820-0",
    journal = "Eur. Phys. J. C",
    volume = "82",
    number = "9",
    pages = "846",
    year = "2022"
}

@article{H1:2010udv,
    author = "Aaron, F. D. and others",
    collaboration = "H1",
    title = "{Inelastic Production of J/psi Mesons in Photoproduction and Deep Inelastic Scattering at HERA}",
    eprint = "1002.0234",
    archivePrefix = "arXiv",
    primaryClass = "hep-ex",
    reportNumber = "DESY-09-225, DESY09-225",
    doi = "10.1140/epjc/s10052-010-1376-5",
    journal = "Eur. Phys. J. C",
    volume = "68",
    pages = "401--420",
    year = "2010"
}

@article{ZEUS:1997wrc,
    author = "Breitweg, J. and others",
    collaboration = "ZEUS",
    title = "{Measurement of inelastic $J/\psi$ photoproduction at HERA}",
    eprint = "hep-ex/9708010",
    archivePrefix = "arXiv",
    reportNumber = "DESY-97-147",
    doi = "10.1007/s002880050583",
    journal = "Z. Phys. C",
    volume = "76",
    pages = "599--612",
    year = "1997"
}

@article{ZEUS:2009Upsilon,
  author         = "{ZEUS Collaboration}",
  title          = "{Exclusive photoproduction of Upsilon mesons at HERA}",
  journal        = "Phys. Lett. B",
  volume         = "680",
  pages          = "4--12",
  year           = "2009",
  doi            = "10.1016/j.physletb.2009.07.066",
  eprint         = "0903.4205",
  archivePrefix  = "arXiv",
  primaryClass   = "hep-ex",
  reportNumber   = "DESY-09-036"
}

@article{Manna:2024ltm,
    author = "Manna, Laboni and Safronov, Anton and Flore, Carlo and Kikola, Daniel and Lansberg, Jean-Philippe and Mattelaer, Olivier",
    title = "{Resolved photoproduction in MadGraph5{\_}aMC@NLO}",
    eprint = "2410.17061",
    archivePrefix = "arXiv",
    primaryClass = "hep-ph",
    doi = "10.22323/1.469.0185",
    journal = "PoS",
    volume = "DIS2024",
    pages = "185",
    year = "2025"
}

@article{Arratia:2020ssx,
    author = "Arratia, Miguel and Makris, Yiannis and Neill, Duff and Ringer, Felix and Sato, Nobuo",
    title = "{Asymmetric jet clustering in deep-inelastic scattering}",
    eprint = "2006.10751",
    archivePrefix = "arXiv",
    primaryClass = "hep-ph",
    reportNumber = "JLAB-THY-20-3209",
    doi = "10.1103/PhysRevD.104.034005",
    journal = "Phys. Rev. D",
    volume = "104",
    number = "3",
    pages = "034005",
    year = "2021"
}

@article{Helenius:2017aqz,
    author = "Helenius, Ilkka",
    editor = "d'Enterria, David and de Roeck, Albert and Mangano, Michelangelo",
    title = "{Photon-photon and photon-hadron processes in Pythia 8}",
    eprint = "1708.09759",
    archivePrefix = "arXiv",
    primaryClass = "hep-ph",
    doi = "10.23727/CERN-Proceedings-2018-001.119",
    journal = "CERN Proc.",
    volume = "1",
    pages = "119",
    year = "2018"
}

@article{Gardner:2023lly,
    author = "Gardner, Simon and Glazier, Derek I. and Livingston, Kenneth and Maneuski, Dzmitry and Sokhan, Daria and Adam, Jaroslav",
    title = "{Pixel-based tracking detectors for a Low Q2 Tagger at EIC -- status report}",
    eprint = "2305.02079",
    archivePrefix = "arXiv",
    primaryClass = "physics.ins-det",
    month = "5",
    year = "2023"
}

@article{AMY:1990F2gamma,
  author        = {Sasaki, T. and others},
  collaboration = {AMY},
  title         = {A measurement of the photon structure function $F_2$},
  journal       = {Phys. Lett. B},
  volume        = {252},
  year          = {1990},
  pages         = {491--498},
}

@article{TPC2gamma:1987F2gamma,
  author        = {Aihara, H. and others},
  collaboration = {TPC/Two-Gamma},
  title         = {An experimental study of the hadronic structure of the photon in
                   $e^+e^- \to e^+e^- + {\rm hadrons}$ at PEP},
  journal       = {Z. Phys. C},
  volume        = {34},
  year          = {1987},
  pages         = {1--13},
}

@article{PLUTO:1981F2gamma,
  author        = {Berger, Ch. and others},
  collaboration = {PLUTO},
  title         = {First measurement of the photon structure function $F_2$},
  journal       = {Phys. Lett. B},
  volume        = {107},
  year          = {1981},
  pages         = {168--172},
}

@article{L3:1999F2gammaQ2,
  author        = {Acciarri, M. and others},
  collaboration = {L3},
  title         = {The $Q^2$ evolution of the hadronic photon structure function $F_2^{\gamma}$ at LEP},
  journal       = {Phys. Lett. B},
  volume        = {447},
  year          = {1999},
  pages         = {147--156},
}

@article{L3:2005F2gamma,
  author        = {Achard, P. and others},
  collaboration = {L3},
  title         = {Measurement of the photon structure function $F_2^{\gamma}$ with the L3 detector at LEP},
  journal       = {Phys. Lett. B},
  volume        = {622},
  year          = {2005},
  pages         = {249--264},
  doi           = {10.1016/j.physletb.2005.07.028},
}

@article{OPAL:1994F2gamma,
  author        = {Akers, R. and others},
  collaboration = {OPAL},
  title         = {Measurement of the photon structure function $F_2^{\gamma}$ in the reaction
                   $e^+e^- \to e^+e^- + {\rm hadrons}$ at LEP},
  journal       = {Z. Phys. C},
  volume        = {61},
  year          = {1994},
  pages         = {199--208},
}

@article{OPAL:2000F2gammaLowX,
  author        = {Abbiendi, G. and others},
  collaboration = {OPAL},
  title         = {Measurement of the low-$x$ behaviour of the photon structure function $F_2^{\gamma}$},
  journal       = {Eur. Phys. J. C},
  volume        = {18},
  year          = {2000},
  pages         = {15--39},
}

@article{OPAL:2002F2gammaLEP2,
  author        = {Abbiendi, G. and others},
  collaboration = {OPAL},
  title         = {Measurement of the hadronic photon structure function $F_2^{\gamma}$ at LEP2},
  journal       = {Phys. Lett. B},
  volume        = {533},
  year          = {2002},
  pages         = {207--222},
}

@article{ALEPH:2003F2gamma,
  author        = {Heister, A. and others},
  collaboration = {ALEPH},
  title         = {Measurement of the hadronic photon structure function $F_2^{\gamma}(x,Q^2)$
                   in two-photon collisions at LEP},
  journal       = {Eur. Phys. J. C},
  volume        = {30},
  year          = {2003},
  pages         = {145--158},
  doi           = {10.1140/epjc/s2003-01291-4},
}

@article{DELPHI:1996F2gamma,
  author        = {Abreu, P. and others},
  collaboration = {DELPHI},
  title         = {A measurement of the photon structure function $F_2^{\gamma}$ at an average
                   $Q^2$ of 12~GeV$^2$},
  journal       = {Z. Phys. C},
  volume        = {69},
  year          = {1996},
  pages         = {223--236},
}

@article{TASSO:1986F2gamma,
  author        = {Althoff, M. and others},
  collaboration = {TASSO},
  title         = {Measurement of the photon structure function $F_2$ at PETRA energies},
  journal       = {Z. Phys. C},
  volume        = {31},
  year          = {1986},
  pages         = {527--538},
}

@article{AMY:1995F2gammaHighQ2,
  author        = {Sahu, S. K. and others},
  collaboration = {AMY},
  title         = {A high-$Q^2$ measurement of the photon structure function $F_2$},
  journal       = {Phys. Lett. B},
  volume        = {346},
  year          = {1995},
  pages         = {208--216},
}

@article{AMY:1997F2gamma,
  author        = {Kojima, T. and others},
  collaboration = {AMY},
  title         = {Measurement of the photon structure function $F_2$ at
                   $\langle Q^2 \rangle \simeq 6.8~\mathrm{GeV}^2$},
  journal       = {Phys. Lett. B},
  volume        = {400},
  year          = {1997},
  pages         = {395--404},
}

@article{TOPAZ:1994F2gamma,
  author        = {Muramatsu, K. and others},
  collaboration = {TOPAZ},
  title         = {Measurement of the photon structure function $F_2$ at TRISTAN},
  journal       = {Phys. Lett. B},
  volume        = {332},
  year          = {1994},
  pages         = {477--486},
}

@article{Meinzinger:2023xuf,
    author = "Meinzinger, Peter and Krauss, Frank",
    title = "{Hadron-level NLO predictions for QCD observables in photo-production at the Electron-Ion Collider}",
    eprint = "2311.14571",
    archivePrefix = "arXiv",
    primaryClass = "hep-ph",
    reportNumber = "IPPP/23/70",
    doi = "10.1103/PhysRevD.109.034037",
    journal = "Phys. Rev. D",
    volume = "109",
    number = "3",
    pages = "034037",
    year = "2024"
}

@article{Bertulani:2024vpt,
    author = "Bertulani, Carlos A. and Francener, Reinaldo and Goncalves, Victor P. and de Souza, Juciene T.",
    title = "{Particle production by {\ensuremath{\gamma}}-{\ensuremath{\gamma}} interactions in future electron-ion colliders}",
    eprint = "2409.00814",
    archivePrefix = "arXiv",
    primaryClass = "hep-ph",
    doi = "10.1103/PhysRevC.111.025201",
    journal = "Phys. Rev. C",
    volume = "111",
    number = "2",
    pages = "025201",
    year = "2025"
}

@article{OPAL2003_Dijet_gammagamma_EPJC31,
  author       = {G. Abbiendi and et al.},
  collaboration= {OPAL},
  title        = {Di-jet production in photon--photon collisions at LEP2},
  journal      = {Eur. Phys. J. C},
  volume       = {31},
  pages        = {307--325},
  year         = {2003}
}

@article{DELPHI2008_Dijet_gammagamma_EPJC57,
  author       = {J. Abdallah and et al.},
  collaboration= {DELPHI},
  title        = {Di-jet production in $\gamma\gamma$ collisions at LEP2},
  journal      = {Eur. Phys. J. C},
  volume       = {57},
  pages        = {525--534},
  year         = {2008}
}

@article{Bassler:1994uq,
    author = "Bassler, Ursula and Bernardi, Gregorio",
    title = "{On the kinematic reconstruction of deep inelastic scattering at HERA: The Sigma method}",
    eprint = "hep-ex/9412004",
    archivePrefix = "arXiv",
    reportNumber = "DESY-94-231",
    doi = "10.1016/0168-9002(95)00173-5",
    journal = "Nucl. Instrum. Meth. A",
    volume = "361",
    pages = "197--208",
    year = "1995"
}

@article{Slominski:2005bw,
    author = "Slominski, W. and Abramowicz, H. and Levy, A.",
    title = "{NLO photon parton parametrization using ee and ep data}",
    eprint = "hep-ph/0504003",
    archivePrefix = "arXiv",
    reportNumber = "TPJU-2-2005, TAUP-2797-05",
    doi = "10.1140/epjc/s2005-02458-7",
    journal = "Eur. Phys. J. C",
    volume = "45",
    pages = "633--641",
    year = "2006"
}

@article{Gordon:1996pm,
    author = "Gordon, L. E. and Storrow, J. K.",
    title = "{New parton distribution functions for the photon}",
    eprint = "hep-ph/9607370",
    archivePrefix = "arXiv",
    reportNumber = "ANL-HEP-PR-96-33, ANL-HEP-PR-96-38, MC-TH-96-16",
    doi = "10.1016/S0550-3213(97)00048-5",
    journal = "Nucl. Phys. B",
    volume = "489",
    pages = "405--426",
    year = "1997"
}

@article{Hoeche:2023gme,
    author = "Hoeche, Stefan and Krauss, Frank and Meinzinger, Peter",
    title = "{Resolved photons in Sherpa}",
    eprint = "2310.18674",
    archivePrefix = "arXiv",
    primaryClass = "hep-ph",
    reportNumber = "FERMILAB-PUB-23-623-T, IPPP/23/59",
    doi = "10.1140/epjc/s10052-024-12551-w",
    journal = "Eur. Phys. J. C",
    volume = "84",
    number = "2",
    pages = "178",
    year = "2024"
}

@article{H1_1998_EffectivePhotonPDF_EPJC1,
  author       = {C. Adloff and et al.},
  collaboration= {H1},
  title        = {Measurement of the inclusive di-jet cross section in photoproduction and determination of an effective parton distribution in the photon},
  journal      = {Eur. Phys. J. C},
  volume       = {1},
  pages        = {97--107},
  year         = {1998},
  doi          = {10.1007/s100520050064}
}

@article{H1_1999_ChargedParticles_GluonPhoton_EPJC10,
  author       = {C. Adloff and et al.},
  collaboration= {H1},
  title        = {Charged particle cross sections in photoproduction and extraction of the gluon density in the photon},
  journal      = {Eur. Phys. J. C},
  volume       = {10},
  pages        = {363--372},
  year         = {1999},
  doi          = {10.1007/s100529900013}
}

@article{H1_2000_VirtualPhotonPDF_EPJC13,
  author       = {C. Adloff and et al.},
  collaboration= {H1},
  title        = {Measurement of dijet cross sections at low $Q^2$ and the extraction of an effective parton density for the virtual photon},
  journal      = {Eur. Phys. J. C},
  volume       = {13},
  pages        = {397--414},
  year         = {2000},
  doi          = {10.1007/s100520050704}
}

@article{ZEUS2002_DijetPhotoprod_arXiv0112029,
  author       = {S. Chekanov and et al.},
  collaboration= {ZEUS},
  title        = {Dijet photoproduction at HERA and the structure of the photon},
  journal      = {Eur. Phys. J. C},
  note         = {arXiv:hep-ex/0112029},
  year         = {2002}
}

@article{Jones:1979wa,
    author = "Jones, L. M. and Sullivan, J. D. and Willen, D. E. and Wyld, H. W.",
    title = "{Photoproduction of {Drell-Yan} Lepton Pairs}",
    reportNumber = "ILL-TH-79-12",
    doi = "10.1103/PhysRevD.20.2749",
    journal = "Phys. Rev. D",
    volume = "20",
    pages = "2749",
    year = "1979",
    note = "[Erratum: Phys.Rev.D 22, 2922 (1980)]"
}

@article{Jaffe:1971we,
    author = "Jaffe, Robert L.",
    title = "{PHOTOPRODUCTION OF MASSIVE MUON PAIRS AT HIGH-ENERGIES}",
    reportNumber = "SLAC-PUB-0913",
    doi = "10.1103/PhysRevD.4.1507",
    journal = "Phys. Rev. D",
    volume = "4",
    pages = "1507",
    year = "1971"
}

@article{Busenitz:1981hr,
    author = "Busenitz, Jerome and Sullivan, J. D.",
    title = "{Photon Structure via the {Drell-Yan} Process}",
    reportNumber = "ILL-TH-81-16",
    doi = "10.1103/PhysRevD.24.1794",
    journal = "Phys. Rev. D",
    volume = "24",
    pages = "1794",
    year = "1981"
}

@article{BrennerMariotto:2013dal,
    author = "Brenner Mariotto, C. and Machado, M. V. T.",
    title = "{Inclusive and exclusive dilepton photoproduction at high energies}",
    eprint = "1303.1439",
    archivePrefix = "arXiv",
    primaryClass = "hep-ph",
    doi = "10.1103/PhysRevD.87.054028",
    journal = "Phys. Rev. D",
    volume = "87",
    number = "5",
    pages = "054028",
    year = "2013"
}

@article{Machado:2008zv,
    author = "Machado, Magno V. T.",
    title = "{Investigating the exclusive protoproduction of dileptons at high energies}",
    eprint = "0805.3144",
    archivePrefix = "arXiv",
    primaryClass = "hep-ph",
    doi = "10.1103/PhysRevD.78.034016",
    journal = "Phys. Rev. D",
    volume = "78",
    pages = "034016",
    year = "2008"
}

@article{Badalian:1990eq,
    author = "Badalian, R. G. and Grabsky, V. O. and Matinyan, Sergei G.",
    title = "{Photoproduction of Drell-Yan leptonic pairs. (In Russian)}",
    journal = "Sov. J. Nucl. Phys.",
    volume = "51",
    pages = "289--293",
    year = "1990"
}

@inproceedings{Bussey:1996vq,
    author = "Bussey, P. J. and Levchenko, B. B. and Shumilin, A.",
    title = "{Prompt photon, Drell-Yan and Bethe-Heitler processes in hard photoproduction}",
    booktitle = "{Workshop on Future Physics at HERA (Preceded by meetings 25-26 Sep 1995 and 7-9 Feb 1996 at DESY)}",
    eprint = "hep-ph/9609273",
    archivePrefix = "arXiv",
    reportNumber = "GLAS-PPE-96-06, INP-MSU-96-26-433",
    pages = "574--579",
    month = "9",
    year = "1996"
}

@article{Witten:1977ju,
    author = "Witten, Edward",
    title = "{Anomalous Cross-Section for Photon - Photon Scattering in Gauge Theories}",
    doi = "10.1016/0550-3213(77)90038-4",
    journal = "Nucl. Phys. B",
    volume = "120",
    pages = "189--202",
    year = "1977"
}

@article{ZEUS:1995xfc,
    author = "Derrick, M. and others",
    collaboration = "ZEUS",
    title = "{Dijet cross-sections in photoproduction at HERA}",
    eprint = "hep-ex/9502008",
    archivePrefix = "arXiv",
    reportNumber = "DESY-95-033",
    doi = "10.1016/0370-2693(95)00275-P",
    journal = "Phys. Lett. B",
    volume = "348",
    pages = "665--680",
    year = "1995"
}

@article{H1:2002apm,
    author = "Adloff, C. and others",
    collaboration = "H1",
    title = "{Measurement of dijet cross-sections in photoproduction at HERA}",
    eprint = "hep-ex/0201006",
    archivePrefix = "arXiv",
    reportNumber = "DESY-01-225",
    doi = "10.1007/s10052-002-1004-0",
    journal = "Eur. Phys. J. C",
    volume = "25",
    pages = "13--23",
    year = "2002"
}

@article{H1:2006rre,
    author = "Aktas, A. and others",
    collaboration = "H1",
    title = "{Photoproduction of dijets with high transverse momenta at HERA}",
    eprint = "hep-ex/0603014",
    archivePrefix = "arXiv",
    reportNumber = "DESY-06-020",
    doi = "10.1016/j.physletb.2006.05.060",
    journal = "Phys. Lett. B",
    volume = "639",
    pages = "21--31",
    year = "2006"
}

@article{ZEUS:2001zoq,
    author = "Chekanov, S. and others",
    collaboration = "ZEUS",
    title = "{Dijet photoproduction at HERA and the structure of the photon}",
    eprint = "hep-ex/0112029",
    archivePrefix = "arXiv",
    reportNumber = "DESY-01-220",
    doi = "10.1007/s100520200936",
    journal = "Eur. Phys. J. C",
    volume = "23",
    pages = "615--631",
    year = "2002"
}

@article{Kang:1979xe,
    author = "Kang, I. and Llewellyn Smith, C. H.",
    title = "{Photoproduction of $\mu$ Pairs and Large $p_T$ Particles in {QCD}}",
    reportNumber = "OXFORD-TP-62/79",
    doi = "10.1016/0550-3213(80)90206-0",
    journal = "Nucl. Phys. B",
    volume = "166",
    pages = "413--428",
    year = "1980"
}

@article{Worden:1974hc,
    author = "Worden, R. P.",
    title = "{Deep Inelastic Structure Functions of the Photon}",
    reportNumber = "RL-74-065",
    doi = "10.1016/0370-2693(74)90151-8",
    journal = "Phys. Lett. B",
    volume = "51",
    pages = "57--62",
    year = "1974"
}

@article{DeWitt:1978wn,
    author = "DeWitt, R. J. and Jones, L. M. and Sullivan, J. D. and Willen, D. E. and Wyld, Jr., H. W.",
    title = "{Anomalous Components of the Photon Structure Functions}",
    reportNumber = "ILL-TH-78-54",
    doi = "10.1103/PhysRevD.19.2046",
    journal = "Phys. Rev. D",
    volume = "19",
    pages = "2046",
    year = "1979",
    note = "[Erratum: Phys.Rev.D 20, 1751 (1979)]"
}

@article{Irving:1980qa,
    author = "Irving, A. C. and Newland, D. B.",
    title = "{PHOTON STRUCTURE FUNCTIONS: ARE THEY MEASURABLE?}",
    reportNumber = "LTH 58",
    doi = "10.1007/BF01427918",
    journal = "Z. Phys. C",
    volume = "6",
    pages = "27",
    year = "1980"
}

@article{Peng:2014hta,
    author = "Peng, Jen-Chieh and Qiu, Jian-Wei",
    title = "{Novel phenomenology of parton distributions from the Drell{\textendash}Yan process}",
    eprint = "1401.0934",
    archivePrefix = "arXiv",
    primaryClass = "hep-ph",
    doi = "10.1016/j.ppnp.2014.01.005",
    journal = "Prog. Part. Nucl. Phys.",
    volume = "76",
    pages = "43--75",
    year = "2014"
}

@article{Berger:2014rva,
    author = "Berger, Ch.",
    title = "{Photon structure function revisited}",
    eprint = "1404.3551",
    archivePrefix = "arXiv",
    primaryClass = "hep-ph",
    doi = "10.4236/jmp.2015.68107",
    journal = "J. Mod. Phys.",
    volume = "6",
    pages = "1023--1043",
    month = "4",
    year = "2014"
}

@inproceedings{Nisius:2009xx,
    author = "Nisius, Richard",
    title = "{Experimental Review of Photon Structure Function Data}",
    booktitle = "{International Conference on the Structure and Interactions of the Photon and 18th International Workshop on Photon-Photon Collisions and International Workshop on High Energy Photon Linear Colliders}",
    eprint = "0907.2782",
    archivePrefix = "arXiv",
    primaryClass = "hep-ex",
    reportNumber = "MPP-2009-131",
    pages = "163--171",
    month = "7",
    year = "2009"
}

@article{NA10:1988ZPC,
  author  = {Guanziroli, M. and others},
  collaboration = {NA10},
  title   = {Angular distributions of muon pairs produced by negative pions on deuterium and tungsten},
  journal = {Z. Phys. C},
  volume  = {37},
  year    = {1988},
  pages   = {545},
  doi     = {10.1007/BF01549713}
}

@article{E615:1989PRD,
  author  = {Conway, J. S. and others},
  collaboration = {E615},
  title   = {Experimental study of muon pairs produced by 252-GeV pions on tungsten},
  journal = {Phys. Rev. D},
  volume  = {39},
  year    = {1989},
  pages   = {92},
  doi     = {10.1103/PhysRevD.39.92}
}

@article{Barry:2018PRL,
  author  = {Barry, P. C. and Sato, N. and Melnitchouk, W. and Ji, C.-R.},
  title   = {First Monte Carlo Global QCD Analysis of Pion Parton Distributions},
  journal = {Phys. Rev. Lett.},
  volume  = {121},
  year    = {2018},
  pages   = {152001},
  doi     = {10.1103/PhysRevLett.121.152001}
}

@article{Novikov:2020xFitter,
  author  = {Novikov, I. and others},
  title   = {Parton distribution functions of the charged pion within the xFitter framework},
  journal = {Phys. Rev. D},
  volume  = {102},
  year    = {2020},
  pages   = {014040},
  doi     = {10.1103/PhysRevD.102.014040}
}

@article{Bourrely:2024KaonPDF,
  author  = {Bourrely, C. and Buccella, F. and Chang, W.-C. and Peng, J.-C.},
  title   = {Extraction of kaon partonic distribution functions from Drell--Yan and $J/\\psi$ production data},
  journal = {Phys. Lett. B},
  volume  = {848},
  year    = {2024},
  pages   = {138395},
  doi     = {10.1016/j.physletb.2023.138395}
}

@article{Peng:2017KaonJpsi,
  author  = {Peng, J.-C. and Chang, W.-C. and Platchkov, S. and Sawada, T.},
  title   = {Valence quark and gluon distributions of kaon from $J/\\psi$ production},
  eprint  = {1711.00839},
  archivePrefix = {arXiv},
  primaryClass = {hep-ph},
  year    = {2017}
}

@article{COMPASS:2024PRL,
  author  = {Alexeev, G. D. and others},
  collaboration = {COMPASS},
  title   = {Final COMPASS Results on the Transverse-Spin-Dependent Azimuthal Asymmetries in the Pion-Induced Drell--Yan Process},
  journal = {Phys. Rev. Lett.},
  volume  = {133},
  year    = {2024},
  pages   = {071902},
  doi     = {10.1103/PhysRevLett.133.071902}
}

@article{H1:2012oud,
    author = "Aaron, F. D. and others",
    collaboration = "H1",
    title = "{Measurement of Beauty and Charm Photoproduction using Semi-muonic Decays in Dijet Events at HERA}",
    eprint = "1205.2495",
    archivePrefix = "arXiv",
    primaryClass = "hep-ex",
    reportNumber = "DESY-12-059",
    doi = "10.1140/epjc/s10052-012-2047-5",
    journal = "Eur. Phys. J. C",
    volume = "72",
    pages = "2047",
    year = "2012"
}

@article{Butterworth:2024hvb,
    author = "Butterworth, Jonathan Mark and Helenius, Ilkka and Castella, Juan Jose Juan and Pattengale, Bradley and Sanjrani, Shahzad and Wing, Matthew",
    title = "{Modelling the underlying event in photon-initiated processes}",
    eprint = "2408.15842",
    archivePrefix = "arXiv",
    primaryClass = "hep-ph",
    doi = "10.21468/SciPostPhys.17.6.158",
    journal = "SciPost Phys.",
    volume = "17",
    number = "6",
    pages = "158",
    year = "2024"
}

@article{ZEUS:2011aa,
    author = "Abramowicz, H. and others",
    collaboration = "ZEUS",
    title = "{Measurement of heavy-quark jet photoproduction at HERA}",
    eprint = "1104.5444",
    archivePrefix = "arXiv",
    primaryClass = "hep-ex",
    reportNumber = "DESY-11-067",
    doi = "10.1140/epjc/s10052-011-1659-5",
    journal = "Eur. Phys. J. C",
    volume = "71",
    pages = "1659",
    year = "2011"
}

@article{PHENIX:2009gyd,
    author = "Adare, A. and others",
    collaboration = "PHENIX",
    title = "{Detailed measurement of the $e^+ e^-$ pair continuum in $p+p$ and Au+Au collisions at $\sqrt{s_{NN}} = 200$ GeV and implications for direct photon production}",
    eprint = "0912.0244",
    archivePrefix = "arXiv",
    primaryClass = "nucl-ex",
    doi = "10.1103/PhysRevC.81.034911",
    journal = "Phys. Rev. C",
    volume = "81",
    pages = "034911",
    year = "2010"
}

@article{ZEUS:2021qzg,
    author = "Abt, I. and others",
    collaboration = "ZEUS",
    title = "{Azimuthal correlations in photoproduction and deep inelastic $ep$ scattering at HERA}",
    eprint = "2106.12377",
    archivePrefix = "arXiv",
    primaryClass = "hep-ex",
    reportNumber = "DESY-21-099",
    doi = "10.1007/JHEP12(2021)102",
    journal = "JHEP",
    volume = "12",
    pages = "102",
    year = "2021"
}

@article{Reimer:2007Review,
  author  = {Reimer, P. E.},
  title   = {Exploring the Partonic Structure of Hadrons through the Drell--Yan Process},
  journal = {J. Phys. G},
  volume  = {34},
  year    = {2007},
  pages   = {S107--S121},
  eprint  = {0704.3621},
  archivePrefix = {arXiv},
  primaryClass = {hep-ex}
}

@article{Brodsky:1971vm,
    author = "Brodsky, Stanley J. and Kinoshita, Toichiro and Terazawa, Hidezumi",
    title = "{Deep inelastic scattering of electrons on a photon target}",
    doi = "10.1103/PhysRevLett.27.280",
    journal = "Phys. Rev. Lett.",
    volume = "27",
    pages = "280--283",
    year = "1971"
}

@article{Sherpa:2019gpd,
    author = "Bothmann, Enrico and others",
    collaboration = "Sherpa",
    title = "{Event Generation with Sherpa 2.2}",
    eprint = "1905.09127",
    archivePrefix = "arXiv",
    primaryClass = "hep-ph",
    reportNumber = "FERMILAB-PUB-19-218-T, SLAC-PUB-17433, IPPP/19/42, MCNET-19-11",
    doi = "10.21468/SciPostPhys.7.3.034",
    journal = "SciPost Phys.",
    volume = "7",
    number = "3",
    pages = "034",
    year = "2019"
}

@article{COMPASS:2023vqt,
    author = "Alexeev, G. D. and others",
    collaboration = "COMPASS",
    title = "{Final COMPASS Results on the Transverse-Spin-Dependent Azimuthal Asymmetries in the Pion-Induced Drell-Yan Process}",
    eprint = "2312.17379",
    archivePrefix = "arXiv",
    primaryClass = "hep-ex",
    reportNumber = "CERN-EP-2023-307",
    doi = "10.1103/PhysRevLett.133.071902",
    journal = "Phys. Rev. Lett.",
    volume = "133",
    number = "7",
    pages = "071902",
    year = "2024"
}

@article{COMPASS2017DYTSA,
  author = {Aghasyan, M. et al. (COMPASS Collaboration)},
  title = {First Measurement of Transverse-Spin-Dependent Azimuthal Asymmetries in the Drell--Yan Process},
  journal = {Phys. Rev. Lett.},
  volume = {119},
  pages = {112002},
  year = {2017},
  doi = {10.1103/PhysRevLett.119.112002},
  eprint = {1704.00488}
}

@article{Walsh:1973mz,
    author = "Walsh, T. F. and Zerwas, Peter M.",
    title = "{Two photon processes in the parton model}",
    reportNumber = "DESY-72-77",
    doi = "10.1016/0370-2693(73)90520-0",
    journal = "Phys. Lett. B",
    volume = "44",
    pages = "195--198",
    year = "1973"
}

@article{Brodsky:1972vv,
    author = "Brodsky, Stanley J. and Close, Francis E. and Gunion, J. F.",
    title = "{Phenomenology of Photon Processes, Vector Dominance and Crucial Tests for Parton Models}",
    reportNumber = "SLAC-PUB-1012",
    doi = "10.1103/PhysRevD.6.177",
    journal = "Phys. Rev. D",
    volume = "6",
    pages = "177",
    year = "1972"
}

@article{Bauer:1977iq,
    author = "Bauer, T. H. and Spital, R. D. and Yennie, D. R. and Pipkin, F. M.",
    title = "{The Hadronic Properties of the Photon in High-Energy Interactions}",
    reportNumber = "PRINT-77-0549 (HARVARD)",
    doi = "10.1103/RevModPhys.50.261",
    journal = "Rev. Mod. Phys.",
    volume = "50",
    pages = "261",
    year = "1978",
    note = "[Erratum: Rev.Mod.Phys. 51, 407 (1979)]"
}

@article{Kroll:1967it,
    author = "Kroll, N. M. and Lee, T. D. and Zumino, B.",
    editor = "Feinberg, G.",
    title = "{Neutral Vector Mesons and the Hadronic Electromagnetic Current}",
    doi = "10.1103/PhysRev.157.1376",
    journal = "Phys. Rev.",
    volume = "157",
    pages = "1376--1399",
    year = "1967"
}

@article{Nisius:1999cv,
    author = "Nisius, Richard",
    title = "{The Photon structure from deep inelastic electron photon scattering}",
    eprint = "hep-ex/9912049",
    archivePrefix = "arXiv",
    doi = "10.1016/S0370-1573(99)00115-5",
    journal = "Phys. Rept.",
    volume = "332",
    pages = "165--317",
    year = "2000"
}

@article{Schuler:1995fk,
    author = "Schuler, Gerhard A. and Sjostrand, Torbjorn",
    title = "{Low and high mass components of the photon distribution functions}",
    eprint = "hep-ph/9503384",
    archivePrefix = "arXiv",
    reportNumber = "CERN-TH-95-62, CERN-TH-95-062, LU-TP-95-6",
    doi = "10.1007/BF01565260",
    journal = "Z. Phys. C",
    volume = "68",
    pages = "607--624",
    year = "1995"
}

@article{Schuler:1996fc,
    author = "Schuler, Gerhard A. and Sjostrand, Torbjorn",
    title = "{Parton distributions of the virtual photon}",
    eprint = "hep-ph/9601282",
    archivePrefix = "arXiv",
    reportNumber = "CERN-TH-96-04, LU-TP-96-2",
    doi = "10.1016/0370-2693(96)00265-1",
    journal = "Phys. Lett. B",
    volume = "376",
    pages = "193--200",
    year = "1996"
}

@article{Ball:2014uwa,
    author = "Ball, Richard D. and others",
    collaboration = "NNPDF",
    title = "{Parton distributions for the LHC Run II}",
    eprint = "1410.8849",
    archivePrefix = "arXiv",
    primaryClass = "hep-ph",
    reportNumber = "EDINBURGH-2014-15, IFUM-1034-FT, CERN-PH-TH-2013-253, OUTP-14-11P, CAVENDISH-HEP-14-11",
    doi = "10.1007/JHEP04(2015)040",
    journal = "JHEP",
    volume = "04",
    pages = "040",
    year = "2015"
}

@article{Jadach:1999vf,
    author = "Jadach, S. and Ward, B. F. L. and Was, Z.",
    title = "{The Precision Monte Carlo event generator K K for two fermion final states in e+ e- collisions}",
    eprint = "hep-ph/9912214",
    archivePrefix = "arXiv",
    reportNumber = "DESY-99-106, CERN-TH-99-235, UTHEP-99-08-01",
    doi = "10.1016/S0010-4655(00)00048-5",
    journal = "Comput. Phys. Commun.",
    volume = "130",
    pages = "260--325",
    year = "2000"
}

@article{Schumann:2007mg,
    author = "Schumann, Steffen and Krauss, Frank",
    title = "{A Parton shower algorithm based on Catani-Seymour dipole factorisation}",
    eprint = "0709.1027",
    archivePrefix = "arXiv",
    primaryClass = "hep-ph",
    reportNumber = "DCPT-07-86, IPPP-07-43",
    doi = "10.1088/1126-6708/2008/03/038",
    journal = "JHEP",
    volume = "03",
    pages = "038",
    year = "2008"
}

@article{Krauss:2001iv,
    author = "Krauss, F. and Kuhn, R. and Soff, G.",
    title = "{AMEGIC++ 1.0: A Matrix element generator in C++}",
    eprint = "hep-ph/0109036",
    archivePrefix = "arXiv",
    reportNumber = "CAVENDISH-HEP-01-11",
    doi = "10.1088/1126-6708/2002/02/044",
    journal = "JHEP",
    volume = "02",
    pages = "044",
    year = "2002"
}

@article{Gleisberg:2008fv,
    author = "Gleisberg, Tanju and Hoeche, Stefan",
    title = "{Comix, a new matrix element generator}",
    eprint = "0808.3674",
    archivePrefix = "arXiv",
    primaryClass = "hep-ph",
    reportNumber = "SLAC-PUB-13232, IPPP-08-31, DCPT-08-62, MCNET-08-08",
    doi = "10.1088/1126-6708/2008/12/039",
    journal = "JHEP",
    volume = "12",
    pages = "039",
    year = "2008"
}

@article{Cascioli:2011va,
    author = "Cascioli, Fabio and Maierhofer, Philipp and Pozzorini, Stefano",
    title = "{Scattering Amplitudes with Open Loops}",
    eprint = "1111.5206",
    archivePrefix = "arXiv",
    primaryClass = "hep-ph",
    reportNumber = "ZU-TH-23-11, LPN11-66",
    doi = "10.1103/PhysRevLett.108.111601",
    journal = "Phys. Rev. Lett.",
    volume = "108",
    pages = "111601",
    year = "2012"
}

@article{Denner:2002ii,
    author = "Denner, Ansgar and Dittmaier, S.",
    title = "{Reduction of one loop tensor five point integrals}",
    eprint = "hep-ph/0212259",
    archivePrefix = "arXiv",
    reportNumber = "MPI-PHT-2002-63, PSI-PR-02-21",
    doi = "10.1016/S0550-3213(03)00184-6",
    journal = "Nucl. Phys. B",
    volume = "658",
    pages = "175--202",
    year = "2003"
}

@article{Denner:2005nn,
    author = "Denner, Ansgar and Dittmaier, S.",
    title = "{Reduction schemes for one-loop tensor integrals}",
    eprint = "hep-ph/0509141",
    archivePrefix = "arXiv",
    reportNumber = "MPP-2005-84, PSI-PR-05-08",
    doi = "10.1016/j.nuclphysb.2005.11.007",
    journal = "Nucl. Phys. B",
    volume = "734",
    pages = "62--115",
    year = "2006"
}

@article{Sullivan:1971kd,
  author       = {Sullivan, J. D.},
  title        = {One-Pion Exchange and Deep-Inelastic Electron-Nucleon Scattering},
  journal      = {Phys. Rev. D},
  volume       = {5},
  pages        = {1732--1737},
  year         = {1972},
  doi          = {10.1103/PhysRevD.5.1732},
  reportNumber = {FERMILAB-PUB-71-008-THY},
}

@article{Denner:2010tr,
    author = "Denner, A. and Dittmaier, S.",
    title = "{Scalar one-loop 4-point integrals}",
    eprint = "1005.2076",
    archivePrefix = "arXiv",
    primaryClass = "hep-ph",
    reportNumber = "FR-PHENO-2010-020, PSI-PR-10-10",
    doi = "10.1016/j.nuclphysb.2010.11.002",
    journal = "Nucl. Phys. B",
    volume = "844",
    pages = "199--242",
    year = "2011"
}

@article{Hoeche:2011fd,
    author = "Hoeche, Stefan and Krauss, Frank and Schonherr, Marek and Siegert, Frank",
    title = "{A critical appraisal of NLO+PS matching methods}",
    eprint = "1111.1220",
    archivePrefix = "arXiv",
    primaryClass = "hep-ph",
    reportNumber = "SLAC-PUB-14661, IPPP-11-67, DCPT-11-134, LPN11-58, FR-PHENO-2011-019, MCNET-11-24",
    doi = "10.1007/JHEP09(2012)049",
    journal = "JHEP",
    volume = "09",
    pages = "049",
    year = "2012"
}

@article{Chahal:2022rid,
    author = "Chahal, Gurpreet Singh and Krauss, Frank",
    title = "{Cluster Hadronisation in Sherpa}",
    eprint = "2203.11385",
    archivePrefix = "arXiv",
    primaryClass = "hep-ph",
    reportNumber = "IPPP/22/14",
    doi = "10.21468/SciPostPhys.13.2.019",
    journal = "SciPost Phys.",
    volume = "13",
    number = "2",
    pages = "019",
    year = "2022"
}

@article{Schonherr:2008av,
    author = "Schonherr, Marek and Krauss, Frank",
    title = "{Soft Photon Radiation in Particle Decays in SHERPA}",
    eprint = "0810.5071",
    archivePrefix = "arXiv",
    primaryClass = "hep-ph",
    reportNumber = "DCPT-07-96, IPPP-07-48, MCNET-08-13",
    doi = "10.1088/1126-6708/2008/12/018",
    journal = "JHEP",
    volume = "12",
    pages = "018",
    year = "2008"
}

@article{Buckley:2019xhk,
    author = {Buckley, Andy and Ilten, Philip and Konstantinov, Dmitri and L{\"o}nnblad, Leif and Monk, James and Pokorski, Witold and Przedzinski, Tomasz and Verbytskyi, Andrii},
    title = "{The HepMC3 event record library for Monte Carlo event generators}",
    eprint = "1912.08005",
    archivePrefix = "arXiv",
    primaryClass = "hep-ph",
    reportNumber = "MPP-2019-258, MCNET-19-27, LU-TP 19-58",
    doi = "10.1016/j.cpc.2020.107310",
    journal = "Comput. Phys. Commun.",
    volume = "260",
    pages = "107310",
    year = "2021"
}

@article{Bothmann:2016nao,
    author = {Bothmann, Enrico and Sch{\"o}nherr, Marek and Schumann, Steffen},
    title = "{Reweighting QCD matrix-element and parton-shower calculations}",
    eprint = "1606.08753",
    archivePrefix = "arXiv",
    primaryClass = "hep-ph",
    reportNumber = "MCNET-16-22, ZU-TH-21-16, ZU--TH--21-16",
    doi = "10.1140/epjc/s10052-016-4430-0",
    journal = "Eur. Phys. J. C",
    volume = "76",
    number = "11",
    pages = "590",
    year = "2016"
}
\end{document}